\documentclass[aps,pra,epsf,superscriptaddress,amsmath,amssymb,amsfonts,twocolumn,showpacs,nofootinbib]{revtex4-1}

\usepackage{epsfig}
\usepackage{dcolumn}
\usepackage{bm}
\usepackage{braket}
\usepackage{amsmath}
\usepackage{mathtools}
\usepackage{graphicx,color,xcolor} 
\usepackage{multirow}
\usepackage{lipsum}
\usepackage{etoolbox}
\usepackage[T1]{fontenc}
\usepackage[colorlinks=true,
            linkcolor=red,
            urlcolor=blue,
            citecolor=blue]{hyperref}

\usepackage[normalem]{ulem} 

\begin{document}

\title{Phases and dynamics of few fermionic impurities  immersed in\\ two-dimensional boson droplets}

\author{Jose Carlos Pelayo$^*$}
\email{jose-pelayo@oist.jp}
\affiliation{Quantum Systems Unit, Okinawa Institute of Science and Technology Graduate University, Okinawa, Japan 904-0495}
\author{ Thom\'{a}s Fogarty}
\affiliation{Quantum Systems Unit, Okinawa Institute of Science and Technology Graduate University, Okinawa, Japan 904-0495}
\author{Thomas Busch}
\affiliation{Quantum Systems Unit, Okinawa Institute of Science and Technology Graduate University, Okinawa, Japan 904-0495}
\author{Simeon I. Mistakidis}
\affiliation{Department of Physics, Missouri University of Science and Technology, Rolla, MO 65409, USA}
\affiliation{ITAMP, Center for Astrophysics $|$ Harvard $\&$ Smithsonian, Cambridge, USA}
\date{\today}

\begin{abstract} 

We unravel the ground state properties and emergent non-equilibrium dynamics of a mixture consisting of a few spin-polarized fermions embedded in a two-dimensional bosonic quantum droplet. For an increasingly attractive droplet-fermion interaction we find a transition from a spatially delocalized fermion configuration to a state where the fermions are highly localized and isolated. This process is accompanied by the rise of induced fermion-fermion interactions mediated by the droplet. Additionally, for increasing attractive droplet-fermion coupling, undulations in the droplet density occur in the vicinity of the fermions manifesting the back-action of the latter. Following interaction quenches from strong attractive to weaker droplet-fermion couplings reveals the spontaneous nucleation of complex excitation patterns in the fermion density such as ring and cross shaped structures. 
These stem from the enhanced interference of the fermions that remain trapped within the droplet, which emulates, to a good degree, an effective potential for the fermions. 
The non-negligible back-action of the droplet manifests itself in the fact that the effective potential predictions are less accurate at the level of the many-body wave function. Our results provide a paradigm for physics beyond the reduced single-component droplet model, unveiling the role of back-action in droplets and the effect of induced mediated interactions.

\end{abstract}

\maketitle

\section{Introduction} 

Quantum droplets in atomic settings are many-body, self-bound states that are nearly incompressible~\cite{droplet_review,bottcher2020new,chomaz2022dipolar}. They are characterized by extremely low densities and are typically around eight orders of magnitude more dilute than their counterparts in  liquid helium~\cite{barranco2006helium}. 
Recently, they have been experimentally realized in single-component~\cite{schmitt2016self,chomaz2022dipolar} and binary~\cite{kirkby2023excitations} dipolar gases but also in Bose mixtures featuring contact interactions~\cite{cabrera2018quantum,cheiney2018bright,fort,semeghini2018self}. 
Their stability originates from the presence of repulsive quantum fluctuations that can be modelled by the perturbative Lee-Huang-Yang (LHY)~\cite{lee1957eigenvalues,larsen1963binary} energy correction which arrests the collapse enforced by mean-field attraction. 
A successful theoretical description of these structures is achieved through the so-called extended Gross-Pitaevskii equation (eGPE), which incorporates the LHY contribution~\cite{petrov2015quantum,Petrov_2016}. The impact of higher-order correlations has also been discussed~\cite{parisi2019liquid,mistakidis2021formation,ota2020beyond,zhang2023density}.

Droplets exhibit a flat-top density for increasing atom number or decreasing  intercomponent attraction. Otherwise, they possess a Gaussian type profile irrespectively of the dimension~\cite{petrov2015quantum,PhysRevA.101.051601,Astrakharchik_2018}, see also Ref.~\cite{holmer2023ground} for solutions at large chemical potentials. 
It has been shown that they can host stable  nonlinear excitations for example in the form of solitary waves~\cite{saqlain2023dragging,katsimiga2023solitary},  vortices~\cite{li2018two,tengstrand2019rotating,gu2023self,yougurt2023vortex} and dispersive shock-waves~\cite{katsimiga2023interactions}. 
These self-bound states can also appear in mixtures with spin-orbit coupling~\cite{tononi2019quantum,PhysRevA.98.023630,gangwar2024spectrum} 
and in Bose-Fermi mixtures~\cite{rakshit2019quantum,rakshit2019self,salasnich2007self}. 
For the latter the competition between an attractive Bose-Fermi coupling and a repulsive Bose-Bose interaction can lead to soliton type structures~\cite{karpiuk2004soliton,karpiuk2006bright}. More recently, it was proposed that higher-order Bose-Fermi interactions are able to support the formation of Bose-Fermi droplets~\cite{rakshit2019quantum,rakshit2019self,salasnich2007self}.

Introducing impurities into such self-bound states can unveil new phenomena related, for instance, to the generation of quasiparticle modes which have been intensively studied in repulsive gases~\cite{massignan2014polarons,schmidt2018universal,mistakidis2023few}, or the existence of induced interactions mediated by the droplet.  
Additionally, impurities provide the possibility to act as probes for the properties of the self-bound configurations, 
and recent investigations have demonstrated that a bosonic impurity embedded in a quasi-one-dimensional Bose-Bose droplet~\cite{abdullaev2020bosonic,sinha2023impurities,bighin2022impurity} features self-localization and a rich excitation spectrum composed of hybrid droplet and impurity modes. On the other hand, it was found that a fermionic impurity immersed in a dipolar droplet allows to tune the bound state character of the latter~\cite{wenzel2018fermionic}. 
The impact of more than a single impurity in a droplet regarding the back-action onto the latter and the phases of the composite system, however, have not yet been explored. 
Here, also induced interactions among the impurities can arise. 
Another interesting prospect is to understand the conditions under which the impurity remains trapped or can escape from the droplet~\cite{sinha2023impurities} when it is dynamically perturbed. 
To address these open questions we consider a few fermionic impurities immersed in a two-dimensional (2D) bosonic droplet with contact interactions.
The ground state and dynamics of this composite system are captured via a set of Schr\"odinger equations for the fermions coupled to an eGPE for the droplet.

We find that the bound character of the ground state is determined by the interplay between the combined mean-field and LHY droplet energy~\cite{Petrov_2016} and the intercomponent one for varying droplet-fermion coupling strengths. 
Also, a larger number of fermions leads to a stronger bound composite system implying that the impurities can manipulate the strength of the ensuing bound state. 
Specifically, a phase-separation~\cite{lous2018probing} between the fermions and the droplet occurs for repulsive intercomponent interactions. 
Importantly, for attractive couplings the fermions delocalize within the droplet and feature a gradual localization for larger attractions, a process reminiscent of the self-pinning transition known in 1D gases~\cite{self-pinning,Keller:23}. 
In this regime, attractive induced interactions among the fermions arise constituting one of our central findings. 
The strength of these induced interactions becomes larger for increasing interspecies attraction and also depends on the number of fermions. 
Simultaneously, a fraction of bosons from the droplet accumulate in the vicinity of the impurities manifesting the back-action of the latter to the droplet for attractive intercomponent couplings.

Turning to the dynamics triggered by a quench of the droplet-fermion interaction from strong to weak attractions, we show that the fermions feature various excitation patterns while remaining trapped within the droplet. 
These excitations originate from interference of the fermion cloud caused by its reflection from the droplet edges. 
The droplet appears to be almost insensitive to the quench exhibiting only weak amplitude breathing motion. 
We also show that the dynamical response of the fermions can be described using an effective approach,  where a static droplet provides an effective potential for the fermions, by demonstrating good quantitative agreement with the predictions of the coupled eGPE model at the density level. However, substantial variations occur at the level of the many-body fermion wave function. 
We note in passing that this effective model is not capable to adequately capture the spatially delocalized fermionic ground state that appears at smaller attractions.

Our manuscript is organized as follows. 
Section~\ref{sec:EGPE} describes the three-component droplet-fermion mixture we consider, the underlying energy functional and set of coupled evolution equations. 
In Section~\ref{sec:GS} we discuss the ground state phases of the droplet-fermion setting and analyze the presence of induced fermion-fermion interactions. 
Section~\ref{sec:dynamics} is devoted to the study of the nonequilibrium dynamics of the mixture following quenches of the intercomponent coupling from strong to weak attraction. 
We conclude and discuss future research directions based on our results in Section~\ref{conclusions}.

\section{Attractively interacting Bose-Fermi mixture}\label{sec:EGPE} 

\subsection{Droplet-fermion setting and assumptions} 

We consider a mixture composed of a few ($N_F$) spin-polarized fermions immersed in a two-component Bose gas. 
 The whole system experiences a strong harmonic confinement of frequency $\omega_z$ in the transversal $z$-direction, such that all energy scales in the $x$-$y$ plane are much smaller than $\hbar\omega_z$. This ensure an effective 
 2D nature of the dynamics 
~\cite{kwon2021spontaneous,makhalov2014ground}, as transversal excitations are essentially frozen out. 
In the two-dimensional plane, both species are confined by a  box potential of length $L_x=L_y \equiv L$ which is chosen large enough  that boundary effects are precluded unless stated otherwise.

To not get lost in the large parameter space, we will assume that the bosonic components have equal repulsive intracomponent 2D $s$-wave scattering lengths, i.e.~$a_{11}=a_{22}\equiv a>0$, and repulsive intercomponent coupling, $a_{12} > 0$. 
In this case,  droplets form in the region where the total energy is negative ~\cite{Petrov_2016,droplet_review}, which appears for densities $n_B < e^{(-2\gamma - 1/2)}\ln{(a_{12}/a)}/2\pi a_{12}a$, with $\gamma$ being  Euler's constant. It is worth noting the difference compared to the case in three dimensions where droplets form when $\delta a^{(3D)}=a_{12}^{(3D)}+\sqrt{a_{11}^{(3D)}a_{22}^{(3D)}}<0$ 
with three-dimensional  scattering lengths $a_{ii}^{(3D)} > 0$ and $a_{12}^{(3D)} < 0$  ~\cite{petrov2015quantum,droplet_review}. 
If we further assume that the bosonic components have the same mass ($m_1=m_2 \equiv m_B$) and atom number ($N_1=N_2 \equiv N_B$), the resulting droplet will be a single one
with $\psi_1=\psi_2 \equiv \psi$ for the two macroscopic bosonic wave functions~\cite{Petrov_2016,hu2022collisional}. For convenience, we also assume that the fermions have the same mass as the bosons ($m_B=m_F \equiv m$). 
However, due to the anti-symmetry of the fermionic many-body wave function, $\Phi(\textbf{r}_1,\textbf{r}_2,...)$, $s$-wave scattering between the individual fermions is forbidden~\cite{pethick2008bose,lewenstein2012ultracold} and they therefore only interact with the bosonic droplet atoms \cite{karpiuk2004soliton}. 

To a good approximation such a droplet-fermion setting can be experimentally realized by a $^{39}\text{K}$-- $^{40}\text{K}$ mixture, where the two bosonic components correspond to two different hyperfine states of $^{39}\text{K}$~\cite{cabrera2018quantum,cheiney2018bright}.

\subsection{Energy functional}

The  energy functional of the  three-component mixture containing the mean-field interactions and the first-order LHY quantum correction can be written as 
\begin{widetext}
\begin{equation}
\begin{split}
    &\frac{E}{V} = \mathcal{E}_{\text{kin}}+\mathcal{E}_{\text{trap}}+\mathcal{E}_{\text{int}},\\
    &\mathcal{E}_{\text{kin}} = \sum_{i=1,2} \frac{\hbar^2}{2m}|\nabla\psi_i(\textbf{r})|^2 + \sum_{i=1}^{N_F}\frac{\hbar^2}{2m}\nabla_i \Phi^*(\textbf{r}_1,\textbf{r}_2,...)\nabla_i \Phi(\textbf{r}_1,\textbf{r}_2,...), \\
    &\mathcal{E}_{\text{trap}} = \sum_{i=1,2} V_{\text{trap}}(\textbf{r})|\psi_i(\textbf{r})|^2 + \sum_{i=1}^{N_F} V_{\text{trap}}(\textbf{r}_i)\Phi^*(\textbf{r}_1,\textbf{r}_2,...)\Phi(\textbf{r}_1,\textbf{r}_2,...), \\
    &\mathcal{E}_{\text{int}} = \underbrace{\sum_{i=1,2}\tilde{g}_{\text{BF}}|\psi_i(\textbf{r})|^2n_F}_\text{$\mathcal{E}_{\text{int}}^{\text{BF}}$} + \underbrace{\frac{\tilde{g}_{BB}}{4}(|\psi_1(\textbf{r})|^2 +  |\psi_2(\textbf{r})|^2)^2\ln\left(\frac{|\psi_1(\textbf{r})|^2+|\psi_2(\textbf{r})|^2}{2en_0}\right)}_\text{$\mathcal{E}_{\text{int}}^{\text{BB}}$}.\label{eq:en_functional}
\end{split}
\end{equation}
\end{widetext} 

Here, $\textbf{r} =(x,y)$, $\tilde{g}_{BB}=8\pi\hbar^2/m\ln^2(a_{12}/a)$ is the intracomponent interaction coefficient for the droplet, and $|\psi_i|^2=n_B$ is the droplet density. The external box potential is given by  $V_{\text{trap}}(x,y) = 0$ for $|x|,|y| \leq L/2$ and $V_{\text{trap}}(x,y) = \infty$ otherwise. 
Moreover, the total fermionic wave function can be expressed as a Hartree product $\Phi = \frac{1}{\sqrt{N_F!}}\sum_i^{N_F!}\text{sgn}(\mathcal{P}_i)\mathcal{P}_i[\phi_1(\textbf{r}_1)...\phi_{N_F}(\textbf{r}_{N_F})]$\cite{karpiuk2004ground}. 
In this expression the $\phi_n$ denote the single-particle fermionic orbitals and $\mathcal{P}_i$ is the permutation operator which exchanges the particle positions within the orbitals. Accordingly, the fermionic density distribution is $n_F = \sum_{n=1}^{N_F}|\phi_n|^2$. 

The first two terms in Eq.~(\ref{eq:en_functional}) represent the standard kinetic ($\mathcal{E}_{\text{kin}}$) and potential ($\mathcal{E}_{\text{trap}}$) energy contributions of the mixture referring to both the droplet components ($i=1,2$) and the fermion subsystems. 
However, the third term $\mathcal{E}_{\text{int}}$ describes the mean-field droplet-fermion interaction energy ($\mathcal{E}_{\text{int}}^{\text{BF}}$) of strength $\tilde{g}_{BF}$, and the combined mean-field and LHY interaction energy terms ($\mathcal{E}_{\text{int}}^{\text{BB}}$) of the bosonic droplet as derived in Ref.~\cite{Petrov_2016}. Here $n_0 = \frac{e^{-2\gamma-3/2}}{2\pi}\frac{\ln(a_{12}/a)}{a_{12}a}$ being the equilibrium droplet density in the thermodynamic limit.

\subsection{Equations of motion}

To identify the ground state phases and monitor the time-evolution of the droplet-fermion mixture we consider the respective system of coupled ($N_F+1$) equations of motion. These correspond to the 2D reduced eGPE describing the droplet and the $N_F$ Hartree-Fock equations of motion for the $\phi_j$ characterizing the time-evolution of the $j$-th fermionic orbital, given by
\begin{subequations}
\begin{align}
&i\frac{\partial\psi}{\partial t} = \bigg[-\frac{\nabla^2}{2} + V_{\text{trap}}+ \nonumber g_{BB}n_B\ln{\left(\frac{n_B}{\sqrt{e}n_0}\right)} \\&~~~~~~~~~~~~~~~~~~~~~~~~~~~~~~+ g_{BF}n_F\bigg]\psi, \label{eq:eGPE_nd}\\
&i\frac{\partial\phi_n}{\partial t} = \left[-\frac{\nabla^2}{2} + V_{\text{trap}}+2g_{BF}n_B\right]\phi_n. \label{eq:Ham_Fermi_nd}
\end{align}\label{coupled system}
\end{subequations}
For generality, the energy, length and time are expressed in terms of $\hbar\omega_z$, $l_{\text{HO}} = \sqrt{\hbar/m\omega_z}$ (transverse oscillator length) and $1/\omega_z$ respectively. 
The interaction parameters are rescaled as $g_{BF}=\tilde{g}_{BF}/(\hbar\omega_z l^2_{HO})$ and $g_{BB}=\tilde{g}_{BB}/(\hbar\omega_z l^2_{HO})$. 

\begin{figure*}[tb] 
\centering  
\includegraphics[width=\textwidth]{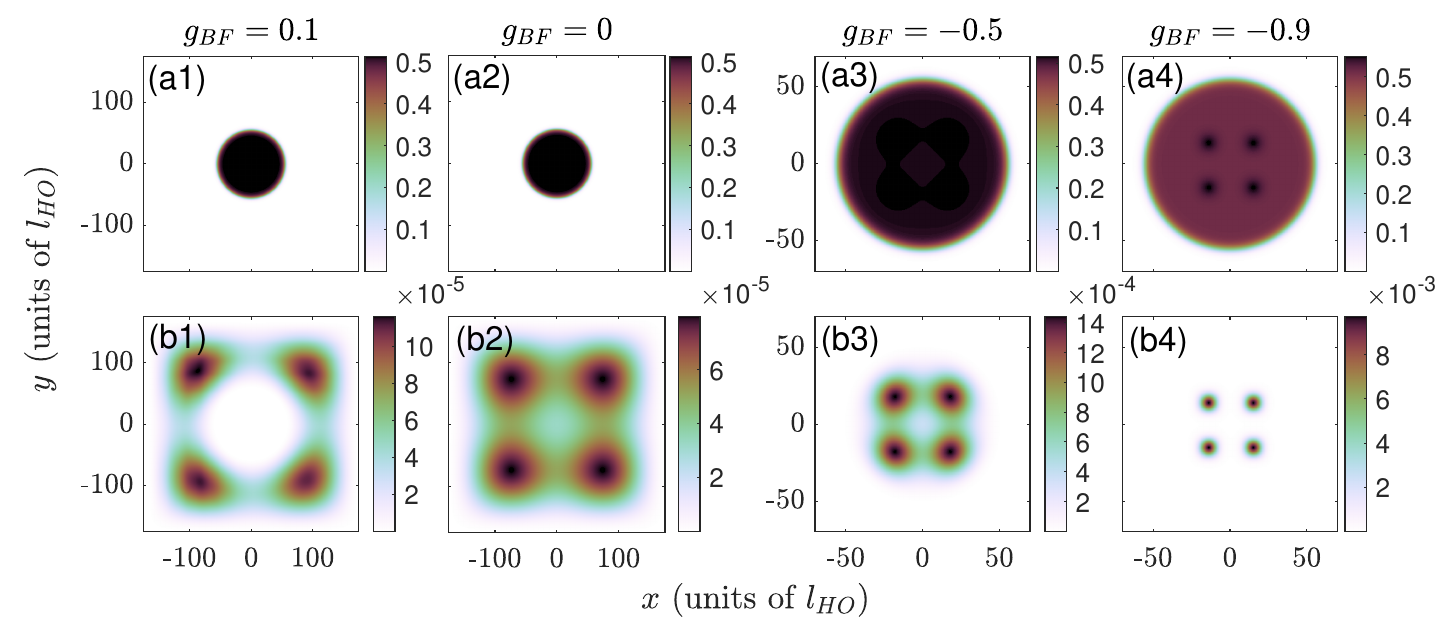}
\caption{(Color online) Ground state densities of (a1)-(a4) the bosonic droplet and (b1)-(b4) the four fermions for decreasing droplet-fermion interaction, $g_{BF}$ (see legends). Phase-separation among the fermions and the droplet occurs for repulsive $g_{BF}$. However, for larger attraction the fermions become gradually more localized and their interparticle distance decreases, while residing inside the droplet. The droplet contains $N_B=5000$ atoms whose interaction is $g_{BB}=0.3112\hbar \omega_z l_{\text{HO}}$. Both components are trapped in a 2D box potential of length $L_x=L_y=L=350l_{\text{HO}}$. The colormap represents the density in units of $l_z^{-2}$. Note that panels (a3)-(b4) are adjusted to a smaller spatial region for better visualization. }
\label{fig:density_N_4}
\end{figure*}

The ground state of the droplet-fermion system is obtained iteratively by evolving Eq.~\eqref{eq:eGPE_nd} in imaginary time using the split-operator method~\cite{weideman1986split}. 
In particular, we first calculate the droplet wave function. Subsequently, we use $n_B$ to diagonalise Eqs.~\eqref{eq:Ham_Fermi_nd} to determine the fermionic wave function. This gives access to $n_F$ and we repeat this scheme  until the energy difference between successive iterations is below a threshold $\sim 10^{-9}$. 
For clarity, it should also be noted that in the above  scheme our initial ansatz are the droplet and fermion ground states at $g_{BF}=0$, see e.g.  Fig.~\ref{fig:density_N_4}(a2), (b2). These states are consecutively used to determine the ones for finite intercomponent coupling $|g_{BF}|>0 $, by adiabatically ramping  $g_{BF}$ in increments of $|\Delta g_{BF}| = 0.01\hbar\omega_z l^2_{HO}$. 
These finite $g_{BF}$ solutions serve as the initial states for the quench-induced dynamics which is monitored through real time 
propagation of the coupled ($N_F+1$) system of Eq.~(\ref{coupled system}).

\section{Ground state configurations }\label{sec:GS} 

Let us first examine the ground state phases of the combined droplet-fermion mixture as a function of the intercomponent interaction $g_{BF}$. 
To get comparable results, we fix the number of bosons in the droplet to be $N_B=5000$ and their intra-component interactions strength as $g_{BB}=0.3112 \hbar \omega_z l_{\text{HO}}$. This corresponds to 2D scattering lengths $a=0.005l_{\text{HO}}$ and $a_{12} = 40l_{\text{HO}}$ such that this subsystem in the absence of fermions forms a 2D flat-top droplet distribution with equilibrium density $n_0 \sim 0.5l_{\text{HO}}^{-2}$, see e.g. Fig.~\ref{fig:density_N_4}(a2). 
In the following we will show how the respective densities, $n_B$ and $n_F$, change as a function of $g_{BF}$. We will start by considering $N_F=4$ impurities, and generalise this number later.

\subsection{Density distributions}

We show representative ground state densities of the droplet (upper panels) and the fermions (lower panels) for different coupling strengths $g_{BF}$ in Fig.~\ref{fig:density_N_4}. The ground state densities for the decoupled case, $g_{BF} = 0$, are depicted in panels (a2) and (b2). Here the droplet can be seen to have a self-bound 2D circularly symmetric, flat-top profile with a peak density of $\text{max}(n_B) \approx n_0$. On the other hand, the fermionic density corresponds to the well-known distribution of non-interacting fermions in a 2D box~\cite{pauli_crystal2}. 
More concretely, the fermions are spatially delocalized and exhibit  a weak spatial overlap with the droplet as their major population is outside of it.

For repulsive droplet-fermion interactions, $g_{BF} > 0$, the components phase-separate (see Fig.~\ref{fig:density_N_4}(a1), (b1)) in order to minimize the interaction energy $E_{\text{int}}^{\text{BF}}$.  Specifically, the fermions lie outside the droplet mainly assembling in four density humps residing at the corners of box and each of them being populated by a single fermion. At the same time the droplet density is largely unchanged. The phase separation process of repulsively interacting Bose-Fermi mixtures has also been previously observed in the case of the bosonic component being in the gaseous phase~\cite{lous2018probing,viverit2000zero,mistakidis2019correlated}. 
It should also be noted that in both the decoupled and the repulsive intercomponent interaction regions, the density overlap between the fermions and the droplet is influenced by finite size effects which  vanish as the box size is enlarged.

Turning to attractive interactions, one can see from Fig.~\ref{fig:density_N_4}(a3)-(b4), that the fermions are located within the droplet, thus maximizing the spatial overlap among the components. 
Here, the droplet-fermion interaction energy $E_{\text{int}}^{\text{BF}}$ is negative [see Fig .~\ref{fig:energy_contribution}(c)] and in fact the energy difference $E_{\text{int}}^{\text{BF}}(g_{BF}<0)-E_{\text{int}}^{\text{BF}}(g_{BF}=0)$ dominates over the remaining energy scales 
[Fig.~\ref{fig:energy_contribution}]. 
This also leads to a fraction of the bosons in the droplet tending towards the vicinity of the fermionic impurities, which cause noticeable undulations in the droplet density especially for increasing intercomponent attraction, see Fig.~\ref{fig:density_N_4}(a4). 
It is worth noting that with increasing $g_{BF}$ attraction the radius of the droplet slightly reduces and the distance between the individual fermions becomes smaller as well. 
This suggests an effective attraction among the fermions mediated by the droplet (see also the discussion below). 
The above demonstrates a transition from a spatially delocalized fermionic distribution to a localized one characterized by well isolated fermions as $g_{BF}$ is tuned to stronger attractive values.
Since the Pauli exclusion principle prevents the fermions from being at the same location, this behaviour resembles the self-pinning transition of a Tonks-Girardeau gas immersed in an one-dimensional Bose gas~\cite{self-pinning}. 

\begin{figure}[tb]
\centering  
\includegraphics[width=\linewidth]{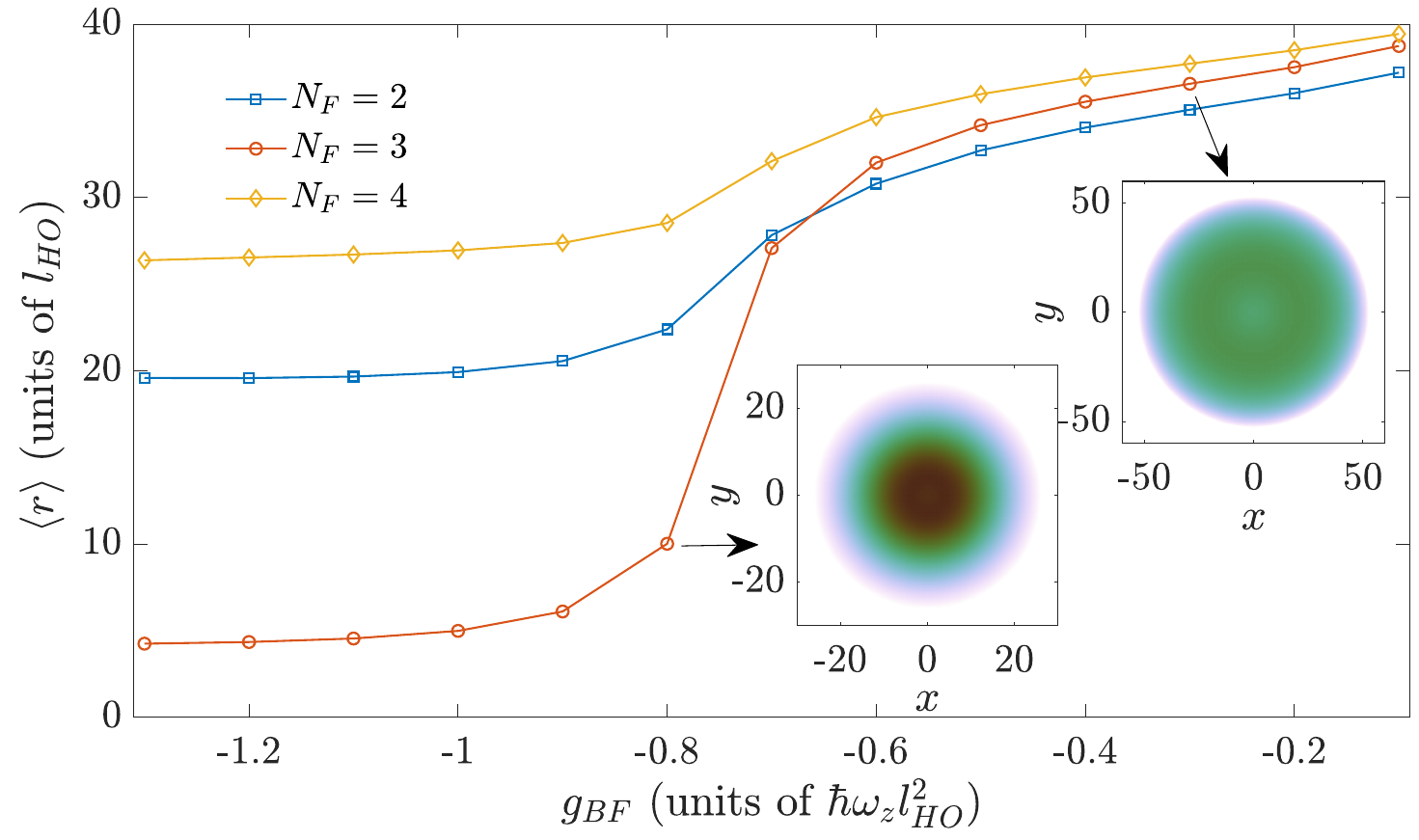}
\caption{(Color online) Relative distance between the fermions with respect to the droplet-fermion interaction strength, $g_{BF}$ for different number of fermions $N_F$ (see legend). It becomes apparent that the relative distance decreases for stronger attractive $g_{BF}$ evincing the existence of mediated attractive interactions among the fermions. Inset depicts the fermionic torus type density  for $N_F=3$ at different $g_{BF}$ (see arrows). Here, the axis are in units of $l_{HO}$ and the colormap depicts the density range in log scale from $n_F = 0 $ to $n_F = 0.01l_z^{-2}$. 
The remaining system parameters are the same as in Fig.~\ref{fig:density_N_4}.}
\label{fig:relative_distance}
\end{figure}

\subsection{Mediated fermion-fermion attraction}

Signatures of induced attractive fermion-fermion interactions can also be seen on the level of the fermion density by computing the relative distance~\cite{distance_measure,distance_measure_1D} between the individual impurities as 
\begin{equation}
    \langle r\rangle = \frac{1}{N_F(N_F - 1)}\int d\textbf{r}_1d\textbf{r}_2|\textbf{r}_1-\textbf{r}_2|\rho^{(2)}(\textbf{r}_1,\textbf{r}_2), 
\end{equation}
where $\rho^{(2)}(\textbf{r}_1,\textbf{r}_2)$ refers to the diagonal  of the fermionic two-body reduced density matrix. 
It determines the probability of simultaneously finding two fermions at positions $\textbf{r}_1$ and $\textbf{r}_2$ respectively~\cite{pethick2008bose,mistakidis2023few}. 
For non-interacting fermions, that we consider here, the diagonal of the two-body reduced density matrix can be expressed in terms of the single-particle orbitals~\cite{density_correlations}, as follows  $\rho^{(2)}(\textbf{r}_1,\textbf{r}_2) = \sum_n |\phi_n(\textbf{r}_1)|^2\sum_m |\phi_m(\textbf{r}_2)|^2 - |\sum_n\phi_n^*(\textbf{r}_1)\phi_n(\textbf{r}_2)|^2 $. 
The relative distance, $\langle r \rangle$, is shown in Fig.~\ref{fig:relative_distance} as a function of $g_{BF}$ for different numbers of fermions. 
Since the fermions move towards the box boundaries for $g_{BF}>0$, due to phase-separation [Fig.~\ref{fig:density_N_4}(b1)], we do not present $\langle r \rangle$ in this interaction regime because it suffers from finite-size effects introduced by the box potential. Let us remark that the relative distance can be  experimentally monitored via an average of in-situ spin-resolved single-shot measurements on the fermionic state~\cite{bergschneider2018spin}.

For $g_{BF} < 0$, we observe that, independently of the number of fermions, $\langle r \rangle$ shows an overall decreasing behavior for stronger attractive droplet-fermion couplings. 
This decreasing trend of $\langle r \rangle$ quantifies the presence of attractive induced fermion-fermion interactions mediated by the droplet. 
Also, the relative reduction of $\langle r \rangle$ with respect to $g_{BF}=0$ provides an estimate for the  strength of induced interactions which can be seen to increase for larger attractions. 
Moreover, at sufficiently large attractions, e.g. here  $g_{BF}<-0.8\hbar \omega_z l_{\text{HO}}$, the relative distance features a saturation tendency thus implying a maximal strength of induced attraction. 
Its worth noting that such a behavior of attractive induced interactions has been reported for both bosonic~\cite{mistakidis2023few,will2021polaron,petkovic2022mediated} and fermionic~\cite{pasek2019induced,baroni2024mediated,mistakidis2019correlated} impurities immersed in a Bose gas and it appears here to equally hold when the medium is a droplet. 
In all cases, $\langle r \rangle$ exhibits a larger rate of decrease at around $g_{BF}\sim-0.7\hbar \omega_z l_{\text{HO}}$ which corresponds to the  interaction region where the onset of the transition from delocalized to isolated fermions occurs, see also Fig.~\ref{fig:density_N_4} for $N_F=4$.  

A notable exception occurs for $N_F=3$, where the fermion distance reduces with a relatively larger rate in the vicinity of $g_{BF}\sim-0.7\hbar \omega_z l_{\text{HO}}$ as compared to $N_F=2,4$. 
However, this does not corresponds to a transition to isolated fermions as in this case $n_F$ can be observed to have a toroidal density profile for weak $g_{BF}$ (see inset of Fig.~\ref{fig:relative_distance}). This shape is maintained  for even stronger attractive $g_{BF}$ where it can be seen to drastically shrink instead of transitioning into a pattern with isolated fermions. 
This torus type fermionic distribution arises due to the underlying closed-shell configuration of the ground state~\cite{pauli_crystal1,pauli_crystal2} and also occurs when the majority component is a weakly-interacting bosonic gas (not shown for brevity).  Here, we confirmed that the same behavior holds in the presence of a droplet. 
Notice also that as in the case of a Bose-Fermi mixture with the bosons being in the gas phase, closed shell configurations are present in the current droplet setting for $N_F =3,6$ fermions in a 2D box that we have checked. However, when $N_F=2$ the aforementioned transition is recovered.

\subsection{Interplay of energy contributions}

\begin{figure*}[tb]
\centering  
\includegraphics[width=0.9\textwidth]{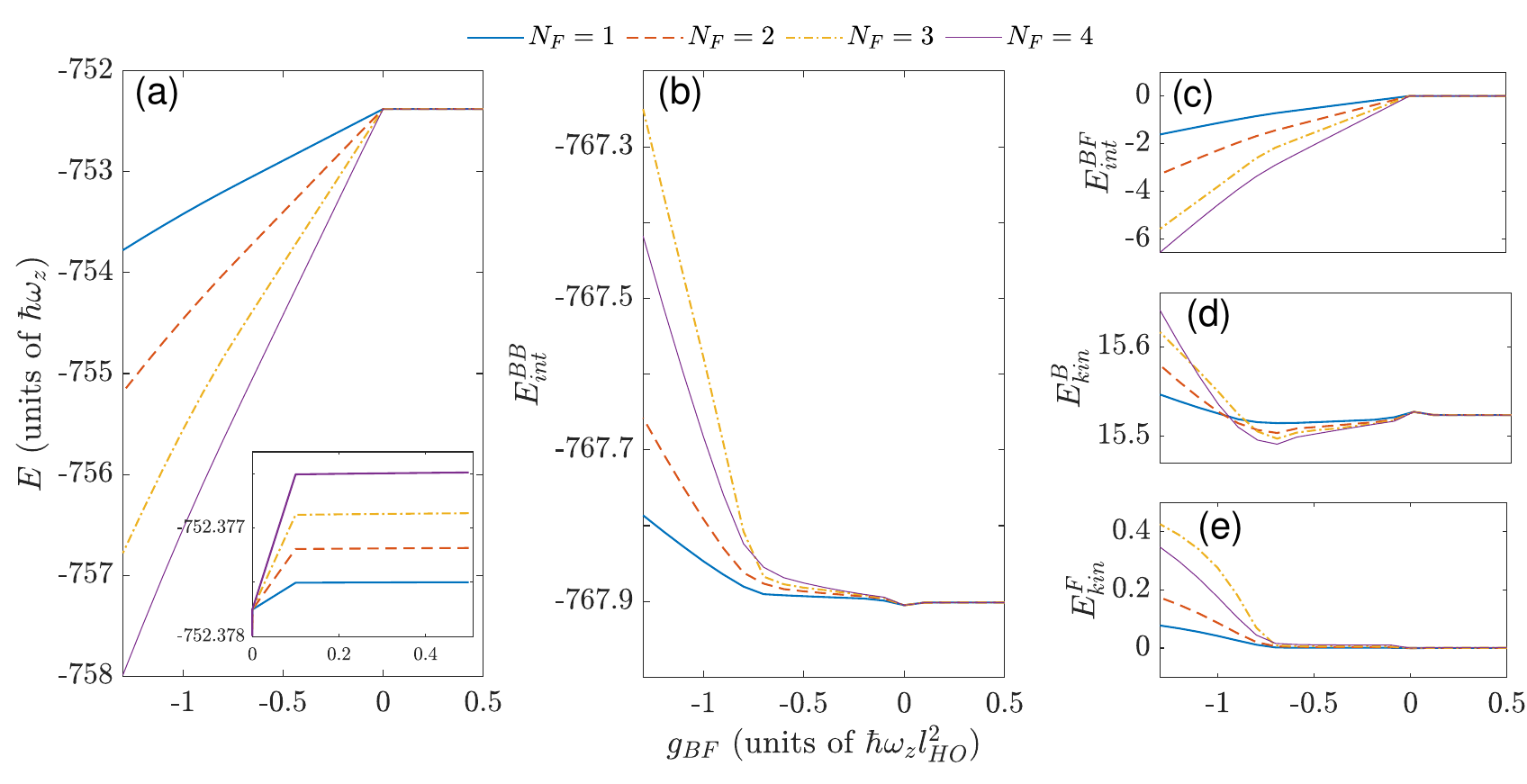}
\caption{(Color online) (a) Ground state energy of the droplet-fermion mixture as a function of $g_{BF}$ for different numbers of fermions (see legend). The increase in the absolute value of the negative energy  for $g_{BF}\leq 0$ indicates the bound character of the system. The inset provides a magnification of the total energy for $g_{BF}>0$. Individual energy contributions of the (b) combined Bose-Bose mean-field and LHY interaction energies, (c) the Bose-Fermi mean-field interaction energy, as well as the kinetic energy of (d) the Bose and (e) the Fermi components. As  can be seen, the interaction energies are negative revealing the origin of the bound state formation. The other system parameters correspond to the ones of Fig.~\ref{fig:density_N_4}.}
\label{fig:energy_contribution}
\end{figure*}

The origin of the above-described droplet-fermion phases lies in the competition between the distinct energy terms given in Eq.~\eqref{eq:en_functional}. 
To better understand this, we show the overall and the  individual energies in Fig.~\ref{fig:energy_contribution} as a function of the intercomponent interaction strength, $g_{BF}$. This first thing to notice is that the overall energy is always negative and dominated by the droplet binding energy, even for positive $g_{BF}$ when no droplet-fermion bound state is present. However, for $g_{BF}\leq 0$, the overall energy $E$ becomes increasingly negative, indicating the creation of the droplet-fermion bound state.
As expected, $E$ becomes more negative for larger attractive $g_{BF}$ and exhibits a hierarchical trend in terms of $N_{F}$. 
Both of these behaviors stem from the interplay of the combined Bose-Bose and LHY interaction energy, $E_{\text{int}}^{\text{BB}}$, and the droplet-fermion interaction energy, $E_{\text{int}}^{\text{BF}}$ shown in Figs.~\ref{fig:energy_contribution}(b) and (c).

Indeed, $E_{\text{int}}^{\text{BB}}<0$ due to the droplet formation, whilst $E_{\text{int}}^{\text{BF}}<0$ for $g_{BF}<0$ and $E_{\text{int}}^{\text{BF}}\approx0$ for $g_{BF}>0$  since phase-separation occurs. 
The most pronounced contribution comes from $E_{\text{int}}^{\text{BB}}$ [Fig.~\ref{fig:energy_contribution}(b)] since it scales as $\sim n_B^2$, while $E_{\text{int}}^{\text{BF}}$ [Fig.~\ref{fig:energy_contribution}(c)] is proportional to $\sim n_B n_F$. 
The latter is also the reason of the aforementioned hierarchical dependence of $E$ since a larger $N_F$ entails an increasingly negative $E_{\text{int}}^{\text{BF}}$. 
This holds also for $g_{BF} > 0$,  where we still observe a tiny increase in $E$ for larger $N_F$ (see inset of Fig.~\ref{fig:energy_contribution}(a)) despite the vanishing overlap among the droplet and the fermions. 
For completeness, we remark that for $N_F = 3$ the slope of $E_{int}^{BB}$ is larger compared to the $N_F = 4$ scenario which can be attributed to the existence of the above-discussed closed shell configurations. 
It should be noted, however, that with respect to $g_{BF}=0$, it is the droplet-fermion relative interaction energy, $E_{\text{int}}^{\text{BF}}(g_{BF} \neq 0)-E_{\text{int}}^{\text{BF}}(g_{BF}=0)$, that is dominant over all other energy terms.

On the other hand, the kinetic energy terms of both the droplet [Fig.~\ref{fig:energy_contribution}(d)], $E_{\text{kin}}^B$, and the fermions [Fig.~\ref{fig:energy_contribution}(d)], $E_{\text{kin}}^F$ remain positive independently of $g_{BF}$. 
In particular, they show a tendency to slightly increase for $g_{BF}<0$ since the fermions are within the droplet and feature induced attraction [see Fig.~\ref{fig:density_N_4}(a3)-(b4)]. 
Moreover, they are almost constant for $g_{BF}>0$ due to phase-separation [see Fig.~\ref{fig:density_N_4}(a1), (b1)].

\begin{figure*}[tb]
\centering  
\includegraphics[width=\textwidth]{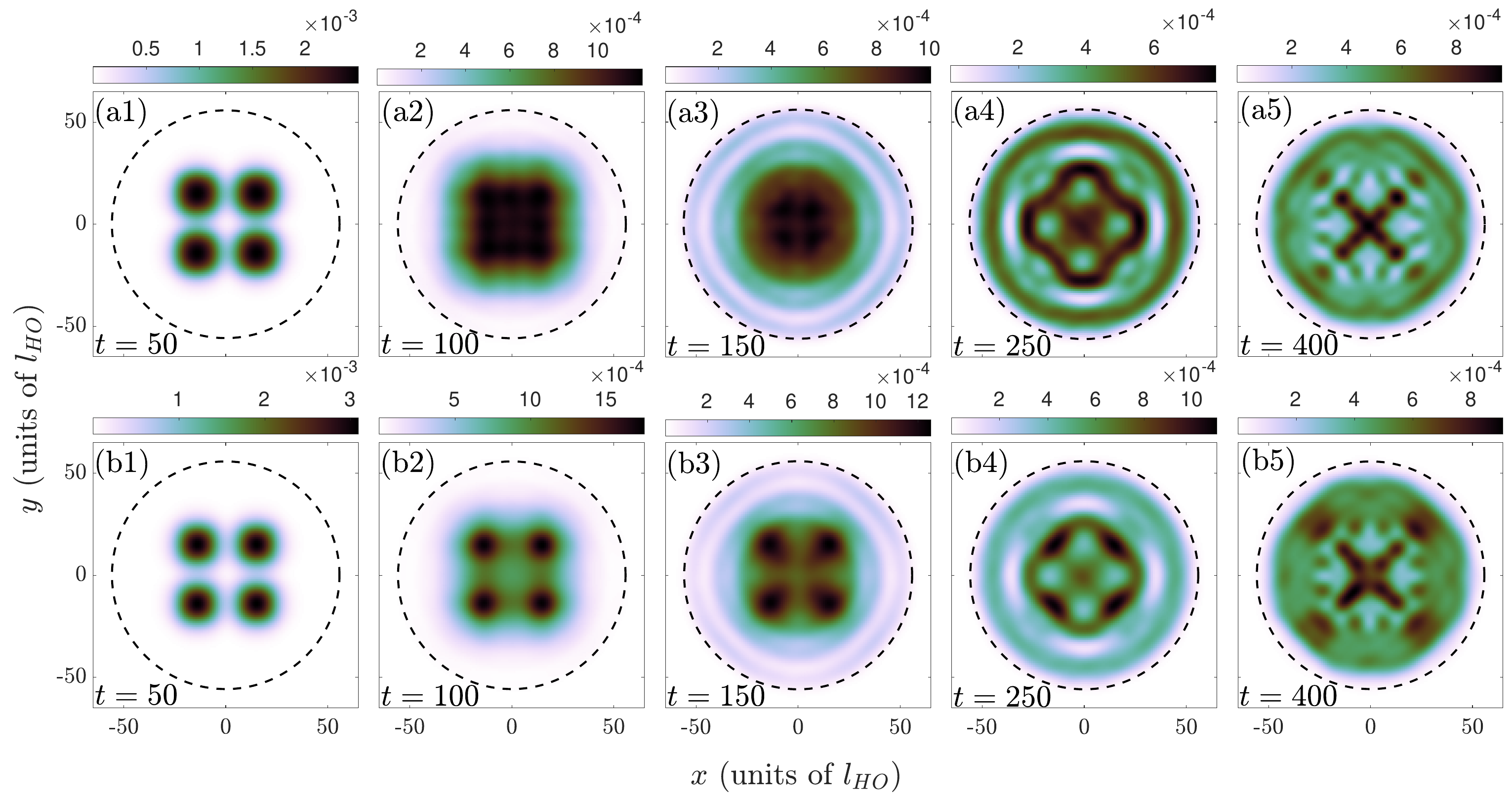}
\caption{(Color online) Snapshots of the fermion density after an interaction quench from $g_{BF_i}=-0.9\hbar \omega_z l_{\text{HO}}$ to $g_{BF_f}=-0.1\hbar \omega_z l_{\text{HO}}$. The fermion dynamics as captured by (a1)-(a5) the complete system [see Eq.~(\ref{coupled system})] and (b1)-(b5) the effective potential approach [see Eq.~(\ref{eq:Ham_Fermi_nd_eff})] is presented. 
Excitation patterns, such as ring and cross type configurations build-upon the fermion density for longer evolution times due to destructive interference of the fermion cloud caused by the droplet edges.
In both cases, the composite system is prepared in its ground state with $g_{BF_i}=-0.9\hbar \omega_z l_{\text{HO}}$, $N_F=4$, $N_B=5000$ and $g_{BB}=0.3112\hbar \omega_z l_{\text{HO}}$ where 
the bosons assemble in a flat-top droplet.  
The black circular dashed lines designate the droplet periphery in the course of the evolution in panels (a1)-(a5) and that of the initial ground state in panels (b1)-(b5).  
The flat-top droplet component experiences weak amplitude  density fluctuations due to the existence of fermions (not shown for brevity). 
The time units correspond to $\omega_z^{-1}$, while the colormap indicates the density in units of $l_z^{-2}$. Qualitative agreement between the two methods is clearly visible. }
\label{fig:dynamics_full_strong_to_weak}
\end{figure*}

\section{Quench induced patterns }\label{sec:dynamics} 

The knowledge of the droplet-fermion ground state phases is a good starting point to study of the dynamical response of the composite system to a sudden perturbation. 
As in our system the inter-component interaction plays a significant role, we will first monitor the system's time-evolution after quenching $g_{BF}$ from the isolated fermion state, e.g. with $g_{BF} = -0.9\hbar \omega_z l_{\text{HO}}$, to the spatially delocalized phase, e.g. for $g_{BF} = -0.1\hbar \omega_z l_{\text{HO}}$.

The emerging 2D fermion density profiles for the above quench are shown in Fig.~\ref{fig:dynamics_full_strong_to_weak}(a1)-(a5). 
Since the postquench interaction is less attractive compared to the prequench one, the originally highly localised fermions [see  Fig.~\ref{fig:density_N_4}(b4)], expand and start noticeable spatially overlapping, see e.g. $t \sim 50\omega_z^{-1}$ in Fig.~\ref{fig:dynamics_full_strong_to_weak}(a1). 
In the course of the evolution, the expansion continues and the major part of the fermion density accumulates at the center forming a square type profile, as depicted for instance in  Fig.~\ref{fig:dynamics_full_strong_to_weak}(a2) at $t \sim 100\omega_z^{-1}$. 
Simultaneously, the tails of the fermion density reach the droplet edges (indicated by the black dashed line in Fig.~\ref{fig:dynamics_full_strong_to_weak})  and are bounced back towards the center where the majority of the fermion cloud resides. 
This behavior suggests that the droplet acts as an effective potential trapping the fermions in its interior and specifically its circular edge emulates a material barrier for the fermions (see also the discussion below). 
The aforementioned reflection of the minority fermion density portion to the center leads to destructive interference with the majority of the fermionic cloud that continuously radially expands outwards.      
As a result, ring shaped structures develop in the fermion density, see Fig.~\ref{fig:dynamics_full_strong_to_weak}(a3) at $t\sim 150\omega_z^{-1}$ where an outer ring is evident in the vicinity of the droplet edge and an inner one closer to the bulk fermion density. 
Notice, however, that these ring structures are shallow, namely their density is not fully dipped and they are not characterized by a $\pi$ phase jump as in the case of ring dark solitons~\cite{kevrekidis2008emergent}.

As time evolves, interference phenomena become pronounced and more complicated excitation patterns appear in the fermion density. 
For instance, at $t\sim 250\omega_z^{-1}$ [Fig.~\ref{fig:dynamics_full_strong_to_weak}(a4)], the outer ring remains close to the droplet edge and it is more prominent while the inner one disappears within the bulk (whose width shrinks) and a deformed rhombik type structure forms. 
Afterwards, the outer ring is also lost and a cross pattern appears in the bulk as a result of the ongoing interference. 
During the whole process, the back-action of the fermions onto the droplet is negligible due to their relatively small density compared to the droplet. 
As such, only weak amplitude density fluctuations occur in the flat-top droplet profile at the vicinity of the fermions and a tiny amplitude breathing mode in the droplet is triggered (not shown). 

\begin{figure}[tb]
\centering  
\includegraphics[width=\linewidth]{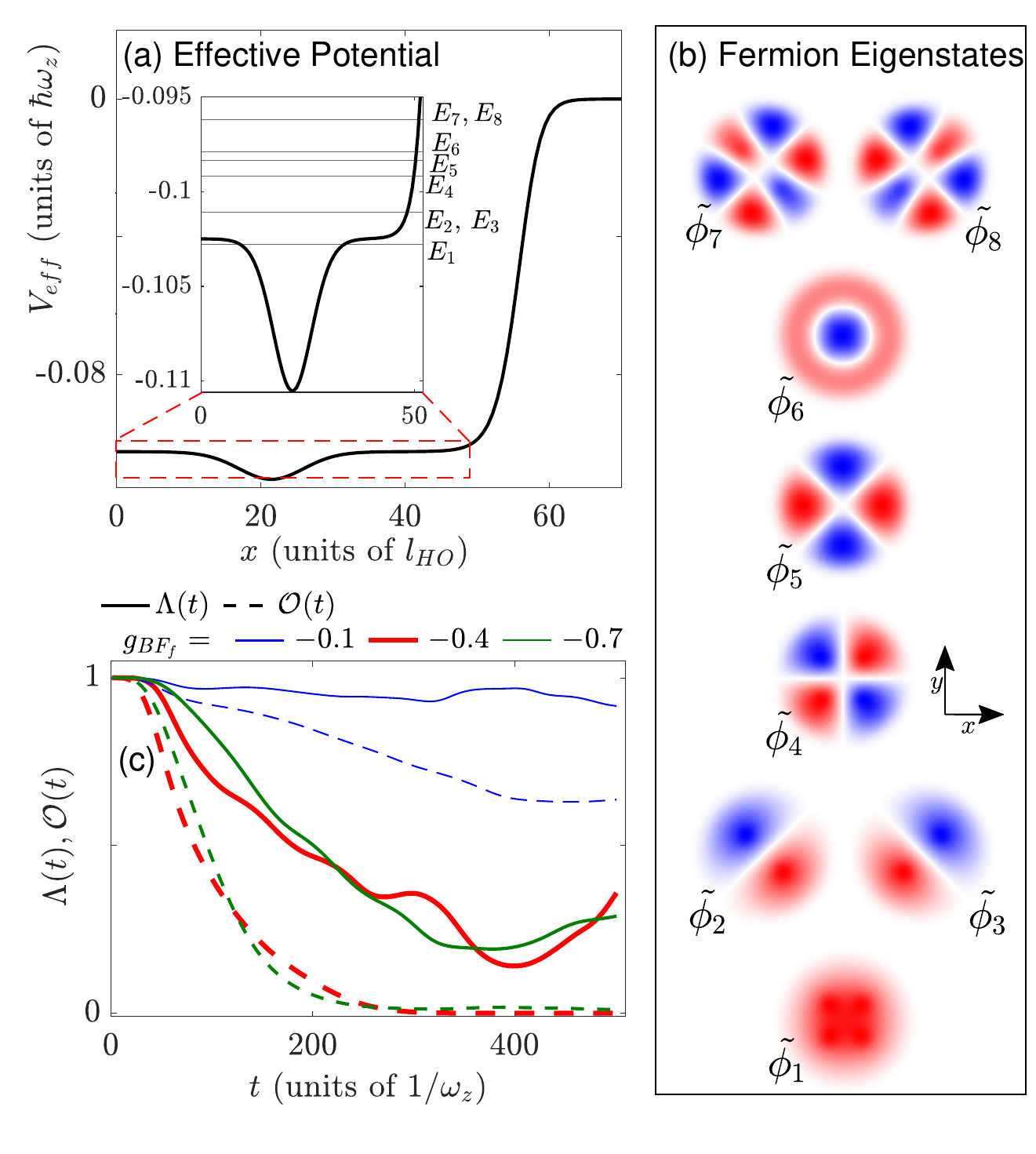}
\caption{(Color online) (a) Profile along $y=x$ (the diagonal) of the effective potential, $V_{\text{eff}}(x)$, created by the ground state distribution of the flat-top droplet for the fermions for $g_{BF_f}=-0.1\hbar \omega_z l_{\text{HO}}$ and $g_{BF_i}=-0.9\hbar \omega_z l_{\text{HO}}$. In practice, the potential is radial. The inset provides a magnification of $V_{\text{eff}}(x)$, in which the horizontal lines represent the eight energetically lowest eigenenergies of the effective potential, with two of them being degenerate (see text). 
(b) The first few low-lying eigenstates of the effective potential. Red (blue) colors refer to negative (positive) values and white to zero. 
(c) The time-evolution of the overlap integral,  $\Lambda(t)$, of the fermion density and the many-body fidelity, $\mathcal{O}(t)$, as predicted from the complete model [Eq.~(\ref{coupled system})] and the effective one [Eq.~(\ref{eq:Ham_Fermi_nd_eff})] for different postquench $g_{BF_f}$ in units of $\hbar \omega_z l_{\text{HO}}$ (see legend). A comparison between $\Lambda(t)$ and $\mathcal{O}(t)$ reveals that on the many-body wavefunction level the former measure underestimates the deviations between the two approaches.}
\label{fig:eff_pot}
\end{figure}

To understand the impact of the droplet on the observed nonequilibrium dynamics we next consider an effective single-component model. 
In this context, the droplet solely acts as a static external potential of the form $V_{\text{eff}} (x,y) = 2g_{BF_f} n_B(t = 0)$ for the fermions. Here, $n_B(t = 0)$ denotes the  droplet ground state density at the initial pre-quench $g_{BF_i}$, while $g_{BF_f}$ is the postquench intercomponent interaction.
A profile in the vicinity of the fermions e.g. at $y=x$ of this effective potential is shown in Fig.~\ref{fig:eff_pot}(a) for $g_{BF_f}=-0.1\hbar \omega_z l_{\text{HO}}$ and a droplet density taken for $g_{BF_i}=-0.9\hbar \omega_z l_{\text{HO}}$. One can see that it has a circular-well shape whose minimum corresponds to the flat-top droplet density. On top of this a dip at the location of each fermion as a result of their back-action to the droplet appears. 

We also show the lowest-lying eigenstates obtained from  diagonalizing  $H_{\text{eff}} = [-\nabla^2 /2 +  V_{\text{eff}} (x,y)]$ for the same parameters as above in Fig.~\ref{fig:eff_pot}(b). The lowest eigenstate $\tilde{\phi}_1$, whose energy lies within the fermion-induced dips in the effective potential, show the characteristic four-humps (hardly visible) on top of a Gaussian-like profile. The next five eigenstates ($\tilde{\phi}_{2}$ to $\tilde{\phi}_{6}$) however closely resemble the eigenstates of non-interacting fermions in a 2D infinite circular well.  Namely, they can be expressed as $\phi_n(r) \sim J_m(k_\nu r)e^{im\theta}$ up to a normalization factor. Here, $r = k_\nu R$ is the $\nu$-th root of the Bessel function of the first kind, $J_m(z)$, for a circular well of radius $R$, and $m \in \{0,\pm 1,\pm 2, ...\}$ is the phase winding. At the same time, $\nu$ also counts one more than the number of radial nodes, that is, for $\nu = 1$, there is no node and we recover the Gaussian-type profile. 
The eigenstate degeneracy is related to the choice of $\pm m$. 
The resemblance of the eigenstates of $H_{\text{eff}}$ to that of the infinite circular well is due to $g_{BF_f}$ being relatively small and thus the fermion-induced density dips are shallow. Additionally, for the higher energy states (e.g. $\tilde{\phi}_{7}$ and $\tilde{\phi}_{8}$) more prominent deviations to the infinite circular well approximation take place where, for instance, regions of positive and negative values are expected to have the same extent.

Under the above-described effective potential  assumption, the time-evolution of the $n$-th fermionic orbital is governed by
\begin{equation}
i\frac{\partial\phi_n(x,y)}{\partial t} = \left[-\frac{\nabla^2}{2} + V_{\text{eff}}(x,y)\right]\phi_n(x,y).  \label{eq:Ham_Fermi_nd_eff}
\end{equation} 
It becomes evident that within the effective approach, a quench of $g_{BF}$ from larger to smaller attractions implies a relatively shallower effective potential. 
The resultant 2D density profiles of the fermions subjected to the aforementioned $V_{\text{eff}}(x,y)$ and obeying Eq.~(\ref{eq:Ham_Fermi_nd_eff}) after a quench from $g_{BF_i}=-0.9\hbar \omega_z l_{\text{HO}}$ to $g_{BF_f}=-0.1\hbar \omega_z l_{\text{HO}}$ are shown in Fig.~\ref{fig:dynamics_full_strong_to_weak}(b1)-(b5) at  times corresponding to the ones where we have shown the densities of the complete model. 
A qualitatively  similar dynamics to the complete model [Eq.~(\ref{coupled system})] can clearly be observed. 

However, certain deviations from the coupled droplet-fermion system exist. 
For example, the density peaks appear to be always larger in the effective system while their locations are fixed to the initial (pre-quenched) location of the fermions even for  relatively long evolution times, for instance at $t\sim250\omega_z^{-1}$ illustrated in Fig.~\ref{fig:dynamics_full_strong_to_weak}(b4). 
Notice also that the merging of the original four fermion density humps and the formation of ring structures are delayed in the effective dynamics, see e.g.  Fig.~\ref{fig:dynamics_full_strong_to_weak}(b3) and (a3).  
Later on, as shown in  Fig.~\ref{fig:dynamics_full_strong_to_weak}(b5), the density spreads out and it still captures the same qualitative features observed within the full approach. 
These differences can therefore be attributed to the neglected density-density droplet-fermion interaction which is also responsible for structural deformations (even small ones) of the droplet during the evolution. 
Notice that this behavior is in line with the ground state one where the effective potential picture becomes gradually invalid for smaller attractions and, for instance, the distribution shown in Fig.~\ref{fig:density_N_4}(b3) can not be recovered.

The degree of deviation in the fermion response between the effective and the extended mean-field models can be tracked either by comparing the fermion densities or their many-body wavefunctions.
Regarding the former case, an estimator is the overlap integral~\cite{jain2011quantum,mistakidis2018correlation}
\begin{equation}
\Lambda(t) = \frac{\left[\int d^2\textbf{r}\,n_F(t)\tilde{n}_F(t)\right]^2}{\int d^2\textbf{r}\, n_F^2(t)\int d^2\textbf{r}\, \tilde{n}_F^2(t)}.  \label{eq:overlap_integral} 
\end{equation} 
The case of $\Lambda(t) = 1$ ($\Lambda(t) = 0$) refers to complete overlap (vanishing overlap), and thus quantifies the difference between the two approaches. Notice that $\tilde{n}_F$ represents the fermion density in the effective system. 
The deviations of the many-body wavefunction can be similarly quantified through the fidelity 
\begin{equation}
   \begin{split}
    \mathcal{O}(t)&=\bigg|\int \Phi(\textbf{r}_1,\textbf{r}_2,\dots,\textbf{r}_{N_F};t)\\
    &\times \Tilde{\Phi}^*(\textbf{r}_1,\textbf{r}_2,\dots,\textbf{r}_{N_F};t)\,d\textbf{r}_1\,d\textbf{r}_2\dots d\textbf{r}_{N_F}\bigg|^2\;.
    \end{split}
\end{equation}
Conveniently, for fermionic systems this can be rewritten in terms of the single-particle orbitals~\cite{fogarty2020orthogonality,goold2011orthogonality}
\begin{align}
\mathcal{O}(t) = |\det[A(t)]|^2,\label{eq:fidelity} 
\end{align}
with the matrix elements $A_{n,m} = \int d^2\textbf{r}\phi_n(\textbf{r},t)\tilde{\phi}^*_m(\textbf{r},t)$ denoting the overlaps between single particle orbitals ${\phi}_n$ and $\tilde{\phi}_m$ of the full and effective model respectively. 
If the fidelity vanishes, $\mathcal{O}(t) \to 0$, the time-evolved many-body fermion wavefunction in the effective approach is orthogonal to the state described by the full model. Meanwhile, for $\mathcal{O}(t) = 1$ the two wavefunctions are identical and the effective approach exactly describes the dynamics. It is important to note that these two  quantities characterise the variations between the two models at different levels.  
Specifically, $\mathcal{O}(t)$ is a more strict measure to gauge the deviations as compared to $\Lambda(t)$ since in the latter case all but one degree of freedom is integrated out.

The time-evolution of both $\Lambda(t)$ and $\mathcal{O}(t)$ for different postquench interactions but fixed initial state is presented in Fig.~\ref{fig:eff_pot}(c). 
As expected, at short times, $0<t<20\omega_z^{-1}$ and independently of $g_{BF_f}$ it holds that $\Lambda(t) \approx 1$ and $\mathcal{O}(t) \approx 1$, indicating an excellent agreement between the two approaches. 
Focusing on $g_{BF_f}=-0.1\hbar \omega_z l_{\text{HO}}$ and longer evolution times, a systematic reduction of $\Lambda(t)$ is observed until $t\approx 320 \omega_z^{-1}$ where the ring and rhombik configurations have been generated as already described above. 
In the time-interval, $320\omega_z^{-1}<t<400\omega_z^{-1}$ where the fermion density features a relatively enhanced spreading within the droplet region [Fig.~\ref{fig:dynamics_full_strong_to_weak}(a5), (b5)], $\Lambda(t)$ naturally increases and afterwards again decreases. 
Overall, $\Lambda(t)$ is not reduced below 0.9 which means that the predicted response between the two methods shows an adequate agreement on the density level.

The many-body fidelity, however, captures larger deviations than the overlap integral in the course of the evolution. More concretely, beyond $t > 20\omega_z^{-1}$ and for $g_{BF_f}=-0.1\hbar \omega_z l_{\text{HO}}$, a gradual decrease of $\mathcal{O}(t)$ during the dynamics takes place reaching a minimum at around  $\mathcal{O}(\gtrsim 400\omega_z^{-1})\sim 0.63$ and subsequently showing a saturation tendency. 
The decreasing behavior of $\mathcal{O}(t)$ evidences that the magnitude of the diagonal elements of $A_{n,m}(t)$ reduces with time indicating an increasing orthogonality trend between the single particle orbitals $\phi_n$ and $\tilde{\phi}_n$.  However, the off-diagonal terms remain comparatively  small in magnitude throughout the evolution.
Additionally, for smaller quench amplitudes, $g_{BF_f} - g_{BF_i}$, e.g. with postquench interactions in the interval $g_{BF_f} = [-0.4,-0.7]\hbar \omega_z l_{\text{HO}}$, a comparatively larger decrease occurs in both $\Lambda(t)$ and $\mathcal{O}(t)$ with the latter being suppressed (i.e. $\mathcal{O}(t)\approx 0$) at longer times. 
The fact that a smaller quench amplitude yields lesser agreement between the two approaches reveals the significant role of the back-action for these postquench attractions.  Namely, for more attractive $g_{BF_f}$ the dynamics is heavily influenced by the droplet back-action to the fermions which is a mechanism not captured by the effective model. 
This is reflected by the finite increase rate of $E_{int}^{BF}$ (in line with the ground state behavior Fig.~\ref{fig:energy_contribution}(c)), within the extended model, thus justifying the deviations between the two approaches. 
Naturally, $\mathcal{O}(t)$ is more sensitive to the interplay of the two-components and thus also the back-action since it encapsulates all degrees-of-freedom of the fermions. Such effects are not adequately captured by $\Lambda(t)$ since the latter solely assesses density modifications.

Finally, we remark that a drastically different response takes place following quenches from weak to strong attractions (not presented for brevity). 
Here, while initially the fermion cloud is spread out in the droplet density, after the quench it shrinks towards the center. This process leads to gradual localization of the density for $t \lesssim 400\omega_z^{-1}$ into four distinct humps resembling the corresponding ground state distribution, see also Fig.~\ref{fig:density_N_4}(b4). 
As time evolves, however, these four density  humps gradually merge at the center forming a  bulk which shapes into different structures such as rectangle or rhombic configuration in the course of the evolution. 
As in the previous quench scenario, the droplet remains to a large extent un-disturbed in the course of the evolution. 
Monitoring the density overlap and the many-body fidelity between the coupled eGPE and the corresponding effective model unveils an overall decreasing trend in both quantities. 
Specifically, they reach a minimum value of  $\Lambda(t/\omega_z^{-1}=329) \approx 0.36$ and $\mathcal{O}(282<t/\omega_z^{-1}<353) \approx 0$ and afterwards they feature a relatively small revival. A similar decreasing behavior followed by a  revival is also observed in $E_{\text{int}}^{BF}(t)$ of the complete system again highlighting the role of the interaction energy in the emergent response.

\section{Conclusions and outlook}\label{conclusions}

We have studied the ground state phases and the corresponding nonequilibrium quantum dynamics of few fermionic impurities embedded in a 2D flat-top bosonic droplet upon variations of the droplet-fermion coupling. 
The bosonic subsystem comprises of two equally populated hyperfine states with the same intracomponent contact interactions, and thus the two-component subsystem can be described by a single-component droplet. 
The droplet lies, in particular, in the flat-top region due to the specific choice of the intercomponent boson interactions that are held fixed. 
The composite system is modelled through a set of $N_F$ mean-field equations for the $N_F$ fermions coupled to a eGPE that takes into account quantum fluctuations for the droplet via the appropriate LHY contribution.

We have shown that the bound character of the composite system is dictated by the combined mean-field and LHY 
as well as the droplet-fermion energy contributions. 
Here, a larger number of fermions results in a more strongly bound system. Different ground state phases of the entire system have been identified depending on the intercomponent interaction strength. 
For repulsive droplet-fermion couplings phase-separation takes place with the fermions residing outside the droplet. Turning to attractive intercomponent interactions it is found that the fermions lie within the droplet and feature a structural deformation for increasing attraction. 
Indeed, they exhibit a spatially delocalized (localized) distribution for weak (strong) attractions. 
Due to the attractive intercomponent coupling a non-negligible fraction of bosons tends towards the location of the impurities. 
This is a consequence of the back-action of the impurities on the droplet in this attractive interaction regime. Interestingly, the aforementioned transition of the fermions for varying droplet-fermion attraction is accompanied by the emergence of attractive induced interactions mediated by the droplet. 
By monitoring the strength of the induced attraction, via the relative distance among the fermions, it is possible to deduce that they are enhanced for larger attraction.

We have also studied the dynamics of the system by inducing a quench of the droplet-fermion coupling from strong to weak attractions. 
After the quench, the isolated fermion state expands towards the droplet edges and  is then reflected back to the center. 
This process triggers the interference of the reflected fermion cloud with the one at the center resulting in peculiar excitation patterns. These include, for instance, the generation of ring-, rhombik-, and cross-shaped configurations. On the other hand, the droplet performs a weak amplitude breathing oscillation and features small density undulations on top of the initial flat-top profile due to the presence of fermions.

Using an effective model where the droplet acts as a static potential for the fermions has been shown to allow for adequate agreement with the coupled eGPE approach regarding the fermion density dynamics. This is not necessarily true for the many-body fidelity of the fermion wave function, where substantial variations among the two methods are evident especially for larger postquench attractions. This inability of the effective model to correctly capture the dynamics at the many-body level can be understood by realising that the effective approach neglects the significant droplet-fermion interaction energy during evolution. 
The effective model also becomes gradually more invalid for smaller attractions since it can not predict the spatially delocalized fermion distributions. 

There are various possible extensions of the work presented here. 
A direct one is to explore using the fermionic impurities to trigger modulational instabilities of the droplet background~\cite{otajonov2022modulational,mithun2020modulational}. 
Another intriguing possibility would be to emulate the respective radiofrequency spectroscopy scheme for the present mixture e.g.~by considering spinor fermionic impurities aiming to establish dressed polaronic states. 
Here, the characterization of the quasi-particle properties such as their residue, effective mass and importantly induced interactions would be  interesting. 
In this context, it would also be valuable to develop an effective model, similar to the ones that have been employed for polarons~\cite{mistakidis2020induced,pasek2019induced}, for quantifying the magnitude and sign of the mediated effective interactions. 
Additionally, the study of induced interactions when two bosonic impurities are embedded within a droplet is an interesting prospect in order to expose their dependence on the different statistics. 
Finally, studying the phase diagram in the crossover towards the particle balance limit of the droplet-fermion setting by systematically increasing the number of fermions and thus enhancing their back-action would be worth pursuing. However, here another approach for the fermions, such as the hydrodynamic one~\cite{rakshit2019quantum}, should be utilized to achieve the description of larger densities.

\section*{Acknowledgements}
This work was supported by the Okinawa Institute of
Science and Technology Graduate University. The authors are grateful for the Scientific Computing and Data Analysis (SCDA) section of the Research Support Division at OIST. The authors thank Tim Keller, Hoshu Hiyane and H. R. Sadeghpour for insightful discussions. T.F. acknowledges support from JSPS KAKENHI Grant No. JP23K03290. T.F. and T.B. are also supported by JST Grant No. JPMJPF2221.

\bibliographystyle{apsrev4-1}
\bibliography{reference}	

\begin{thebibliography}{73}%
\makeatletter
\providecommand \@ifxundefined [1]{%
 \@ifx{#1\undefined}
}%
\providecommand \@ifnum [1]{%
 \ifnum #1\expandafter \@firstoftwo
 \else \expandafter \@secondoftwo
 \fi
}%
\providecommand \@ifx [1]{%
 \ifx #1\expandafter \@firstoftwo
 \else \expandafter \@secondoftwo
 \fi
}%
\providecommand \natexlab [1]{#1}%
\providecommand \enquote  [1]{``#1''}%
\providecommand \bibnamefont  [1]{#1}%
\providecommand \bibfnamefont [1]{#1}%
\providecommand \citenamefont [1]{#1}%
\providecommand \href@noop [0]{\@secondoftwo}%
\providecommand \href [0]{\begingroup \@sanitize@url \@href}%
\providecommand \@href[1]{\@@startlink{#1}\@@href}%
\providecommand \@@href[1]{\endgroup#1\@@endlink}%
\providecommand \@sanitize@url [0]{\catcode `\\12\catcode `\$12\catcode `\&12\catcode `\#12\catcode `\^12\catcode `\_12\catcode `\%12\relax}%
\providecommand \@@startlink[1]{}%
\providecommand \@@endlink[0]{}%
\providecommand \url  [0]{\begingroup\@sanitize@url \@url }%
\providecommand \@url [1]{\endgroup\@href {#1}{\urlprefix }}%
\providecommand \urlprefix  [0]{URL }%
\providecommand \Eprint [0]{\href }%
\providecommand \doibase [0]{http://dx.doi.org/}%
\providecommand \selectlanguage [0]{\@gobble}%
\providecommand \bibinfo  [0]{\@secondoftwo}%
\providecommand \bibfield  [0]{\@secondoftwo}%
\providecommand \translation [1]{[#1]}%
\providecommand \BibitemOpen [0]{}%
\providecommand \bibitemStop [0]{}%
\providecommand \bibitemNoStop [0]{.\EOS\space}%
\providecommand \EOS [0]{\spacefactor3000\relax}%
\providecommand \BibitemShut  [1]{\csname bibitem#1\endcsname}%
\let\auto@bib@innerbib\@empty
\bibitem [{\citenamefont {Luo}\ \emph {et~al.}(2021)\citenamefont {Luo}, \citenamefont {Pang}, \citenamefont {Liu}, \citenamefont {Li},\ and\ \citenamefont {Malomed}}]{droplet_review}%
  \BibitemOpen
  \bibfield  {author} {\bibinfo {author} {\bibfnamefont {Z.-H.}\ \bibnamefont {Luo}}, \bibinfo {author} {\bibfnamefont {W.}~\bibnamefont {Pang}}, \bibinfo {author} {\bibfnamefont {B.}~\bibnamefont {Liu}}, \bibinfo {author} {\bibfnamefont {Y.-Y.}\ \bibnamefont {Li}}, \ and\ \bibinfo {author} {\bibfnamefont {B.~A.}\ \bibnamefont {Malomed}},\ }\href {\doibase https://doi.org/10.1007/s11467-020-1020-2} {\bibfield  {journal} {\bibinfo  {journal} {Front. Phys.}\ }\textbf {\bibinfo {volume} {16}},\ \bibinfo {pages} {1} (\bibinfo {year} {2021})}\BibitemShut {NoStop}%
\bibitem [{\citenamefont {B{\"o}ttcher}\ \emph {et~al.}(2020)\citenamefont {B{\"o}ttcher}, \citenamefont {Schmidt}, \citenamefont {Hertkorn}, \citenamefont {Ng}, \citenamefont {Graham}, \citenamefont {Guo}, \citenamefont {Langen},\ and\ \citenamefont {Pfau}}]{bottcher2020new}%
  \BibitemOpen
  \bibfield  {author} {\bibinfo {author} {\bibfnamefont {F.}~\bibnamefont {B{\"o}ttcher}}, \bibinfo {author} {\bibfnamefont {J.-N.}\ \bibnamefont {Schmidt}}, \bibinfo {author} {\bibfnamefont {J.}~\bibnamefont {Hertkorn}}, \bibinfo {author} {\bibfnamefont {K.~S.}\ \bibnamefont {Ng}}, \bibinfo {author} {\bibfnamefont {S.~D.}\ \bibnamefont {Graham}}, \bibinfo {author} {\bibfnamefont {M.}~\bibnamefont {Guo}}, \bibinfo {author} {\bibfnamefont {T.}~\bibnamefont {Langen}}, \ and\ \bibinfo {author} {\bibfnamefont {T.}~\bibnamefont {Pfau}},\ }\href {\doibase 10.1088/1361-6633/abc9ab} {\bibfield  {journal} {\bibinfo  {journal} {Rep. Progr. Phys.}\ }\textbf {\bibinfo {volume} {84}},\ \bibinfo {pages} {012403} (\bibinfo {year} {2020})}\BibitemShut {NoStop}%
\bibitem [{\citenamefont {Chomaz}\ \emph {et~al.}(2022)\citenamefont {Chomaz}, \citenamefont {Ferrier-Barbut}, \citenamefont {Ferlaino}, \citenamefont {Laburthe-Tolra}, \citenamefont {Lev},\ and\ \citenamefont {Pfau}}]{chomaz2022dipolar}%
  \BibitemOpen
  \bibfield  {author} {\bibinfo {author} {\bibfnamefont {L.}~\bibnamefont {Chomaz}}, \bibinfo {author} {\bibfnamefont {I.}~\bibnamefont {Ferrier-Barbut}}, \bibinfo {author} {\bibfnamefont {F.}~\bibnamefont {Ferlaino}}, \bibinfo {author} {\bibfnamefont {B.}~\bibnamefont {Laburthe-Tolra}}, \bibinfo {author} {\bibfnamefont {B.~L.}\ \bibnamefont {Lev}}, \ and\ \bibinfo {author} {\bibfnamefont {T.}~\bibnamefont {Pfau}},\ }\href {\doibase 10.1088/1361-6633/aca814} {\bibfield  {journal} {\bibinfo  {journal} {Rep. Progr. Phys.}\ } (\bibinfo {year} {2022}),\ 10.1088/1361-6633/aca814}\BibitemShut {NoStop}%
\bibitem [{\citenamefont {Barranco}\ \emph {et~al.}(2006)\citenamefont {Barranco}, \citenamefont {Guardiola}, \citenamefont {Hern{\'a}ndez}, \citenamefont {Mayol}, \citenamefont {Navarro},\ and\ \citenamefont {Pi}}]{barranco2006helium}%
  \BibitemOpen
  \bibfield  {author} {\bibinfo {author} {\bibfnamefont {M.}~\bibnamefont {Barranco}}, \bibinfo {author} {\bibfnamefont {R.}~\bibnamefont {Guardiola}}, \bibinfo {author} {\bibfnamefont {S.}~\bibnamefont {Hern{\'a}ndez}}, \bibinfo {author} {\bibfnamefont {R.}~\bibnamefont {Mayol}}, \bibinfo {author} {\bibfnamefont {J.}~\bibnamefont {Navarro}}, \ and\ \bibinfo {author} {\bibfnamefont {M.}~\bibnamefont {Pi}},\ }\href {\doibase https://doi.org/10.1007/s10909-005-9267-0} {\bibfield  {journal} {\bibinfo  {journal} {J. L. Temp. Phys.}\ }\textbf {\bibinfo {volume} {142}},\ \bibinfo {pages} {1} (\bibinfo {year} {2006})}\BibitemShut {NoStop}%
\bibitem [{\citenamefont {Schmitt}\ \emph {et~al.}(2016)\citenamefont {Schmitt}, \citenamefont {Wenzel}, \citenamefont {B{\"o}ttcher}, \citenamefont {Ferrier-Barbut},\ and\ \citenamefont {Pfau}}]{schmitt2016self}%
  \BibitemOpen
  \bibfield  {author} {\bibinfo {author} {\bibfnamefont {M.}~\bibnamefont {Schmitt}}, \bibinfo {author} {\bibfnamefont {M.}~\bibnamefont {Wenzel}}, \bibinfo {author} {\bibfnamefont {F.}~\bibnamefont {B{\"o}ttcher}}, \bibinfo {author} {\bibfnamefont {I.}~\bibnamefont {Ferrier-Barbut}}, \ and\ \bibinfo {author} {\bibfnamefont {T.}~\bibnamefont {Pfau}},\ }\href {\doibase https://doi.org/10.1038/nature20126} {\bibfield  {journal} {\bibinfo  {journal} {Nature}\ }\textbf {\bibinfo {volume} {539}},\ \bibinfo {pages} {259} (\bibinfo {year} {2016})}\BibitemShut {NoStop}%
\bibitem [{\citenamefont {Kirkby}\ \emph {et~al.}(2023)\citenamefont {Kirkby}, \citenamefont {Lee}, \citenamefont {Baillie}, \citenamefont {Bland}, \citenamefont {Ferlaino}, \citenamefont {Blakie},\ and\ \citenamefont {Bisset}}]{kirkby2023excitations}%
  \BibitemOpen
  \bibfield  {author} {\bibinfo {author} {\bibfnamefont {W.}~\bibnamefont {Kirkby}}, \bibinfo {author} {\bibfnamefont {A.-C.}\ \bibnamefont {Lee}}, \bibinfo {author} {\bibfnamefont {D.}~\bibnamefont {Baillie}}, \bibinfo {author} {\bibfnamefont {T.}~\bibnamefont {Bland}}, \bibinfo {author} {\bibfnamefont {F.}~\bibnamefont {Ferlaino}}, \bibinfo {author} {\bibfnamefont {P.~B.}\ \bibnamefont {Blakie}}, \ and\ \bibinfo {author} {\bibfnamefont {R.~N.}\ \bibnamefont {Bisset}},\ }\href {\doibase https://doi.org/10.48550/arXiv.2312.03390} {\bibfield  {journal} {\bibinfo  {journal} {arXiv:2312.03390}\ } (\bibinfo {year} {2023}),\ https://doi.org/10.48550/arXiv.2312.03390}\BibitemShut {NoStop}%
\bibitem [{\citenamefont {Cabrera}\ \emph {et~al.}(2018)\citenamefont {Cabrera}, \citenamefont {Tanzi}, \citenamefont {Sanz}, \citenamefont {Naylor}, \citenamefont {Thomas}, \citenamefont {Cheiney},\ and\ \citenamefont {Tarruell}}]{cabrera2018quantum}%
  \BibitemOpen
  \bibfield  {author} {\bibinfo {author} {\bibfnamefont {C.~R.}\ \bibnamefont {Cabrera}}, \bibinfo {author} {\bibfnamefont {L.}~\bibnamefont {Tanzi}}, \bibinfo {author} {\bibfnamefont {J.}~\bibnamefont {Sanz}}, \bibinfo {author} {\bibfnamefont {B.}~\bibnamefont {Naylor}}, \bibinfo {author} {\bibfnamefont {P.}~\bibnamefont {Thomas}}, \bibinfo {author} {\bibfnamefont {P.}~\bibnamefont {Cheiney}}, \ and\ \bibinfo {author} {\bibfnamefont {L.}~\bibnamefont {Tarruell}},\ }\href {\doibase 10.1126/science.aao5686} {\bibfield  {journal} {\bibinfo  {journal} {Science}\ }\textbf {\bibinfo {volume} {359}},\ \bibinfo {pages} {301} (\bibinfo {year} {2018})}\BibitemShut {NoStop}%
\bibitem [{\citenamefont {Cheiney}\ \emph {et~al.}(2018)\citenamefont {Cheiney}, \citenamefont {Cabrera}, \citenamefont {Sanz}, \citenamefont {Naylor}, \citenamefont {Tanzi},\ and\ \citenamefont {Tarruell}}]{cheiney2018bright}%
  \BibitemOpen
  \bibfield  {author} {\bibinfo {author} {\bibfnamefont {P.}~\bibnamefont {Cheiney}}, \bibinfo {author} {\bibfnamefont {C.~R.}\ \bibnamefont {Cabrera}}, \bibinfo {author} {\bibfnamefont {J.}~\bibnamefont {Sanz}}, \bibinfo {author} {\bibfnamefont {B.}~\bibnamefont {Naylor}}, \bibinfo {author} {\bibfnamefont {L.}~\bibnamefont {Tanzi}}, \ and\ \bibinfo {author} {\bibfnamefont {L.}~\bibnamefont {Tarruell}},\ }\href {\doibase 10.1103/PhysRevLett.120.135301} {\bibfield  {journal} {\bibinfo  {journal} {Phys. Rev. Lett.}\ }\textbf {\bibinfo {volume} {120}},\ \bibinfo {pages} {135301} (\bibinfo {year} {2018})}\BibitemShut {NoStop}%
\bibitem [{\citenamefont {D'Errico}\ \emph {et~al.}(2019)\citenamefont {D'Errico}, \citenamefont {Burchianti}, \citenamefont {Prevedelli}, \citenamefont {Salasnich}, \citenamefont {Ancilotto}, \citenamefont {Modugno}, \citenamefont {Minardi},\ and\ \citenamefont {Fort}}]{fort}%
  \BibitemOpen
  \bibfield  {author} {\bibinfo {author} {\bibfnamefont {C.}~\bibnamefont {D'Errico}}, \bibinfo {author} {\bibfnamefont {A.}~\bibnamefont {Burchianti}}, \bibinfo {author} {\bibfnamefont {M.}~\bibnamefont {Prevedelli}}, \bibinfo {author} {\bibfnamefont {L.}~\bibnamefont {Salasnich}}, \bibinfo {author} {\bibfnamefont {F.}~\bibnamefont {Ancilotto}}, \bibinfo {author} {\bibfnamefont {M.}~\bibnamefont {Modugno}}, \bibinfo {author} {\bibfnamefont {F.}~\bibnamefont {Minardi}}, \ and\ \bibinfo {author} {\bibfnamefont {C.}~\bibnamefont {Fort}},\ }\href {\doibase 10.1103/PhysRevResearch.1.033155} {\bibfield  {journal} {\bibinfo  {journal} {Phys. Rev. Research}\ }\textbf {\bibinfo {volume} {1}},\ \bibinfo {pages} {033155} (\bibinfo {year} {2019})}\BibitemShut {NoStop}%
\bibitem [{\citenamefont {Semeghini}\ \emph {et~al.}(2018)\citenamefont {Semeghini}, \citenamefont {Ferioli}, \citenamefont {Masi}, \citenamefont {Mazzinghi}, \citenamefont {Wolswijk}, \citenamefont {Minardi}, \citenamefont {Modugno}, \citenamefont {Modugno}, \citenamefont {Inguscio},\ and\ \citenamefont {Fattori}}]{semeghini2018self}%
  \BibitemOpen
  \bibfield  {author} {\bibinfo {author} {\bibfnamefont {G.}~\bibnamefont {Semeghini}}, \bibinfo {author} {\bibfnamefont {G.}~\bibnamefont {Ferioli}}, \bibinfo {author} {\bibfnamefont {L.}~\bibnamefont {Masi}}, \bibinfo {author} {\bibfnamefont {C.}~\bibnamefont {Mazzinghi}}, \bibinfo {author} {\bibfnamefont {L.}~\bibnamefont {Wolswijk}}, \bibinfo {author} {\bibfnamefont {F.}~\bibnamefont {Minardi}}, \bibinfo {author} {\bibfnamefont {M.}~\bibnamefont {Modugno}}, \bibinfo {author} {\bibfnamefont {G.}~\bibnamefont {Modugno}}, \bibinfo {author} {\bibfnamefont {M.}~\bibnamefont {Inguscio}}, \ and\ \bibinfo {author} {\bibfnamefont {M.}~\bibnamefont {Fattori}},\ }\href {\doibase https://doi.org/10.1103/PhysRevLett.120.235301} {\bibfield  {journal} {\bibinfo  {journal} {Phys. Rev. Lett.}\ }\textbf {\bibinfo {volume} {120}},\ \bibinfo {pages} {235301} (\bibinfo {year} {2018})}\BibitemShut {NoStop}%
\bibitem [{\citenamefont {Lee}\ \emph {et~al.}(1957)\citenamefont {Lee}, \citenamefont {Huang},\ and\ \citenamefont {Yang}}]{lee1957eigenvalues}%
  \BibitemOpen
  \bibfield  {author} {\bibinfo {author} {\bibfnamefont {T.~D.}\ \bibnamefont {Lee}}, \bibinfo {author} {\bibfnamefont {K.}~\bibnamefont {Huang}}, \ and\ \bibinfo {author} {\bibfnamefont {C.~N.}\ \bibnamefont {Yang}},\ }\href {\doibase https://doi.org/10.1103/PhysRev.106.1135} {\bibfield  {journal} {\bibinfo  {journal} {Phys. Rev.}\ }\textbf {\bibinfo {volume} {106}},\ \bibinfo {pages} {1135} (\bibinfo {year} {1957})}\BibitemShut {NoStop}%
\bibitem [{\citenamefont {Larsen}(1963)}]{larsen1963binary}%
  \BibitemOpen
  \bibfield  {author} {\bibinfo {author} {\bibfnamefont {D.~M.}\ \bibnamefont {Larsen}},\ }\href {\doibase https://doi.org/10.1016/0003-4916(63)90066-6} {\bibfield  {journal} {\bibinfo  {journal} {Annals of Physics (New York)(US)}\ }\textbf {\bibinfo {volume} {24}} (\bibinfo {year} {1963}),\ https://doi.org/10.1016/0003-4916(63)90066-6}\BibitemShut {NoStop}%
\bibitem [{\citenamefont {Petrov}(2015)}]{petrov2015quantum}%
  \BibitemOpen
  \bibfield  {author} {\bibinfo {author} {\bibfnamefont {D.~S.}\ \bibnamefont {Petrov}},\ }\href {\doibase https://doi.org/10.1103/PhysRevLett.115.155302} {\bibfield  {journal} {\bibinfo  {journal} {Phys. Rev. Lett.}\ }\textbf {\bibinfo {volume} {115}},\ \bibinfo {pages} {155302} (\bibinfo {year} {2015})}\BibitemShut {NoStop}%
\bibitem [{\citenamefont {Petrov}\ and\ \citenamefont {Astrakharchik}(2016)}]{Petrov_2016}%
  \BibitemOpen
  \bibfield  {author} {\bibinfo {author} {\bibfnamefont {D.~S.}\ \bibnamefont {Petrov}}\ and\ \bibinfo {author} {\bibfnamefont {G.~E.}\ \bibnamefont {Astrakharchik}},\ }\href {\doibase 10.1103/PhysRevLett.117.100401} {\bibfield  {journal} {\bibinfo  {journal} {Phys. Rev. Lett.}\ }\textbf {\bibinfo {volume} {117}},\ \bibinfo {pages} {100401} (\bibinfo {year} {2016})}\BibitemShut {NoStop}%
\bibitem [{\citenamefont {Parisi}\ \emph {et~al.}(2019)\citenamefont {Parisi}, \citenamefont {Astrakharchik},\ and\ \citenamefont {Giorgini}}]{parisi2019liquid}%
  \BibitemOpen
  \bibfield  {author} {\bibinfo {author} {\bibfnamefont {L.}~\bibnamefont {Parisi}}, \bibinfo {author} {\bibfnamefont {G.~E.}\ \bibnamefont {Astrakharchik}}, \ and\ \bibinfo {author} {\bibfnamefont {S.}~\bibnamefont {Giorgini}},\ }\href {\doibase 10.1103/PhysRevLett.122.105302} {\bibfield  {journal} {\bibinfo  {journal} {Phys. Rev. Lett.}\ }\textbf {\bibinfo {volume} {122}},\ \bibinfo {pages} {105302} (\bibinfo {year} {2019})}\BibitemShut {NoStop}%
\bibitem [{\citenamefont {Mistakidis}\ \emph {et~al.}(2021)\citenamefont {Mistakidis}, \citenamefont {Mithun}, \citenamefont {Kevrekidis}, \citenamefont {Sadeghpour},\ and\ \citenamefont {Schmelcher}}]{mistakidis2021formation}%
  \BibitemOpen
  \bibfield  {author} {\bibinfo {author} {\bibfnamefont {S.~I.}\ \bibnamefont {Mistakidis}}, \bibinfo {author} {\bibfnamefont {T.}~\bibnamefont {Mithun}}, \bibinfo {author} {\bibfnamefont {P.~G.}\ \bibnamefont {Kevrekidis}}, \bibinfo {author} {\bibfnamefont {H.~R.}\ \bibnamefont {Sadeghpour}}, \ and\ \bibinfo {author} {\bibfnamefont {P.}~\bibnamefont {Schmelcher}},\ }\href {\doibase https://doi.org/10.1103/PhysRevResearch.3.043128} {\bibfield  {journal} {\bibinfo  {journal} {Phys. Rev. Research}\ }\textbf {\bibinfo {volume} {3}},\ \bibinfo {pages} {043128} (\bibinfo {year} {2021})}\BibitemShut {NoStop}%
\bibitem [{\citenamefont {Ota}\ and\ \citenamefont {Astrakharchik}(2020)}]{ota2020beyond}%
  \BibitemOpen
  \bibfield  {author} {\bibinfo {author} {\bibfnamefont {M.}~\bibnamefont {Ota}}\ and\ \bibinfo {author} {\bibfnamefont {G.}~\bibnamefont {Astrakharchik}},\ }\href {\doibase 10.21468/SciPostPhys.9.2.020} {\bibfield  {journal} {\bibinfo  {journal} {SciPost Phys.}\ }\textbf {\bibinfo {volume} {9}},\ \bibinfo {pages} {020} (\bibinfo {year} {2020})}\BibitemShut {NoStop}%
\bibitem [{\citenamefont {Zhang}\ and\ \citenamefont {Yin}(2023)}]{zhang2023density}%
  \BibitemOpen
  \bibfield  {author} {\bibinfo {author} {\bibfnamefont {F.}~\bibnamefont {Zhang}}\ and\ \bibinfo {author} {\bibfnamefont {L.}~\bibnamefont {Yin}},\ }\href {\doibase https://doi.org/10.48550/arXiv.2306.00254} {\bibfield  {journal} {\bibinfo  {journal} {arXiv:2306.00254}\ } (\bibinfo {year} {2023}),\ https://doi.org/10.48550/arXiv.2306.00254}\BibitemShut {NoStop}%
\bibitem [{\citenamefont {Tylutki}\ \emph {et~al.}(2020)\citenamefont {Tylutki}, \citenamefont {Astrakharchik}, \citenamefont {Malomed},\ and\ \citenamefont {Petrov}}]{PhysRevA.101.051601}%
  \BibitemOpen
  \bibfield  {author} {\bibinfo {author} {\bibfnamefont {M.}~\bibnamefont {Tylutki}}, \bibinfo {author} {\bibfnamefont {G.~E.}\ \bibnamefont {Astrakharchik}}, \bibinfo {author} {\bibfnamefont {B.~A.}\ \bibnamefont {Malomed}}, \ and\ \bibinfo {author} {\bibfnamefont {D.~S.}\ \bibnamefont {Petrov}},\ }\href {\doibase 10.1103/PhysRevA.101.051601} {\bibfield  {journal} {\bibinfo  {journal} {Phys. Rev. A}\ }\textbf {\bibinfo {volume} {101}},\ \bibinfo {pages} {051601(R)} (\bibinfo {year} {2020})}\BibitemShut {NoStop}%
\bibitem [{\citenamefont {Astrakharchik}\ and\ \citenamefont {Malomed}(2018)}]{Astrakharchik_2018}%
  \BibitemOpen
  \bibfield  {author} {\bibinfo {author} {\bibfnamefont {G.~E.}\ \bibnamefont {Astrakharchik}}\ and\ \bibinfo {author} {\bibfnamefont {B.~A.}\ \bibnamefont {Malomed}},\ }\href {\doibase 10.1103/PhysRevA.98.013631} {\bibfield  {journal} {\bibinfo  {journal} {Phys. Rev. A}\ }\textbf {\bibinfo {volume} {98}},\ \bibinfo {pages} {013631} (\bibinfo {year} {2018})}\BibitemShut {NoStop}%
\bibitem [{\citenamefont {Holmer}\ \emph {et~al.}(2023)\citenamefont {Holmer}, \citenamefont {Zhang},\ and\ \citenamefont {Kevrekidis}}]{holmer2023ground}%
  \BibitemOpen
  \bibfield  {author} {\bibinfo {author} {\bibfnamefont {J.}~\bibnamefont {Holmer}}, \bibinfo {author} {\bibfnamefont {K.~Z.}\ \bibnamefont {Zhang}}, \ and\ \bibinfo {author} {\bibfnamefont {P.~G.}\ \bibnamefont {Kevrekidis}},\ }\href {\doibase https://doi.org/10.48550/arXiv.2401.00213} {\bibfield  {journal} {\bibinfo  {journal} {arXiv:2401.00213}\ } (\bibinfo {year} {2023}),\ https://doi.org/10.48550/arXiv.2401.00213}\BibitemShut {NoStop}%
\bibitem [{\citenamefont {Saqlain}\ \emph {et~al.}(2023)\citenamefont {Saqlain}, \citenamefont {Mithun}, \citenamefont {Carretero-Gonz{\'a}lez},\ and\ \citenamefont {Kevrekidis}}]{saqlain2023dragging}%
  \BibitemOpen
  \bibfield  {author} {\bibinfo {author} {\bibfnamefont {S.}~\bibnamefont {Saqlain}}, \bibinfo {author} {\bibfnamefont {T.}~\bibnamefont {Mithun}}, \bibinfo {author} {\bibfnamefont {R.}~\bibnamefont {Carretero-Gonz{\'a}lez}}, \ and\ \bibinfo {author} {\bibfnamefont {P.~G.}\ \bibnamefont {Kevrekidis}},\ }\href {\doibase https://doi.org/10.1103/PhysRevA.107.033310} {\bibfield  {journal} {\bibinfo  {journal} {Phys. Rev. A}\ }\textbf {\bibinfo {volume} {107}},\ \bibinfo {pages} {033310} (\bibinfo {year} {2023})}\BibitemShut {NoStop}%
\bibitem [{\citenamefont {Katsimiga}\ \emph {et~al.}(2023{\natexlab{a}})\citenamefont {Katsimiga}, \citenamefont {Mistakidis}, \citenamefont {Koutsokostas}, \citenamefont {Frantzeskakis}, \citenamefont {Carretero-Gonz{\'a}lez},\ and\ \citenamefont {Kevrekidis}}]{katsimiga2023solitary}%
  \BibitemOpen
  \bibfield  {author} {\bibinfo {author} {\bibfnamefont {G.~C.}\ \bibnamefont {Katsimiga}}, \bibinfo {author} {\bibfnamefont {S.~I.}\ \bibnamefont {Mistakidis}}, \bibinfo {author} {\bibfnamefont {G.~N.}\ \bibnamefont {Koutsokostas}}, \bibinfo {author} {\bibfnamefont {D.~J.}\ \bibnamefont {Frantzeskakis}}, \bibinfo {author} {\bibfnamefont {R.}~\bibnamefont {Carretero-Gonz{\'a}lez}}, \ and\ \bibinfo {author} {\bibfnamefont {P.~G.}\ \bibnamefont {Kevrekidis}},\ }\href {\doibase https://doi.org/10.1103/PhysRevA.107.063308} {\bibfield  {journal} {\bibinfo  {journal} {Phys. Rev. A}\ }\textbf {\bibinfo {volume} {107}},\ \bibinfo {pages} {063308} (\bibinfo {year} {2023}{\natexlab{a}})}\BibitemShut {NoStop}%
\bibitem [{\citenamefont {Li}\ \emph {et~al.}(2018)\citenamefont {Li}, \citenamefont {Chen}, \citenamefont {Luo}, \citenamefont {Huang}, \citenamefont {Tan}, \citenamefont {Pang},\ and\ \citenamefont {Malomed}}]{li2018two}%
  \BibitemOpen
  \bibfield  {author} {\bibinfo {author} {\bibfnamefont {Y.}~\bibnamefont {Li}}, \bibinfo {author} {\bibfnamefont {Z.}~\bibnamefont {Chen}}, \bibinfo {author} {\bibfnamefont {Z.}~\bibnamefont {Luo}}, \bibinfo {author} {\bibfnamefont {C.}~\bibnamefont {Huang}}, \bibinfo {author} {\bibfnamefont {H.}~\bibnamefont {Tan}}, \bibinfo {author} {\bibfnamefont {W.}~\bibnamefont {Pang}}, \ and\ \bibinfo {author} {\bibfnamefont {B.~A.}\ \bibnamefont {Malomed}},\ }\href {\doibase https://doi.org/10.1103/PhysRevA.98.063602} {\bibfield  {journal} {\bibinfo  {journal} {Phys. Rev. A}\ }\textbf {\bibinfo {volume} {98}},\ \bibinfo {pages} {063602} (\bibinfo {year} {2018})}\BibitemShut {NoStop}%
\bibitem [{\citenamefont {Tengstrand}\ \emph {et~al.}(2019)\citenamefont {Tengstrand}, \citenamefont {St{\"u}rmer}, \citenamefont {Karabulut},\ and\ \citenamefont {Reimann}}]{tengstrand2019rotating}%
  \BibitemOpen
  \bibfield  {author} {\bibinfo {author} {\bibfnamefont {M.~N.}\ \bibnamefont {Tengstrand}}, \bibinfo {author} {\bibfnamefont {P.}~\bibnamefont {St{\"u}rmer}}, \bibinfo {author} {\bibfnamefont {E.}~\bibnamefont {Karabulut}}, \ and\ \bibinfo {author} {\bibfnamefont {S.~M.}\ \bibnamefont {Reimann}},\ }\href {\doibase https://doi.org/10.1103/PhysRevLett.123.160405} {\bibfield  {journal} {\bibinfo  {journal} {Phys. Rev. Lett.}\ }\textbf {\bibinfo {volume} {123}},\ \bibinfo {pages} {160405} (\bibinfo {year} {2019})}\BibitemShut {NoStop}%
\bibitem [{\citenamefont {Gu}\ and\ \citenamefont {Cui}(2023)}]{gu2023self}%
  \BibitemOpen
  \bibfield  {author} {\bibinfo {author} {\bibfnamefont {Q.}~\bibnamefont {Gu}}\ and\ \bibinfo {author} {\bibfnamefont {X.}~\bibnamefont {Cui}},\ }\href {\doibase https://doi.org/10.1103/PhysRevA.108.063302} {\bibfield  {journal} {\bibinfo  {journal} {Phys. Rev. A}\ }\textbf {\bibinfo {volume} {108}},\ \bibinfo {pages} {063302} (\bibinfo {year} {2023})}\BibitemShut {NoStop}%
\bibitem [{\citenamefont {Yo{\u{g}}urt}\ \emph {et~al.}(2023)\citenamefont {Yo{\u{g}}urt}, \citenamefont {Tanyeri}, \citenamefont {Kele{\c{s}}},\ and\ \citenamefont {Oktel}}]{yougurt2023vortex}%
  \BibitemOpen
  \bibfield  {author} {\bibinfo {author} {\bibfnamefont {T.~A.}\ \bibnamefont {Yo{\u{g}}urt}}, \bibinfo {author} {\bibfnamefont {U.}~\bibnamefont {Tanyeri}}, \bibinfo {author} {\bibfnamefont {A.}~\bibnamefont {Kele{\c{s}}}}, \ and\ \bibinfo {author} {\bibfnamefont {M.}~\bibnamefont {Oktel}},\ }\href {\doibase https://doi.org/10.1103/PhysRevA.108.033315} {\bibfield  {journal} {\bibinfo  {journal} {Phys. Rev. A}\ }\textbf {\bibinfo {volume} {108}},\ \bibinfo {pages} {033315} (\bibinfo {year} {2023})}\BibitemShut {NoStop}%
\bibitem [{\citenamefont {Katsimiga}\ \emph {et~al.}(2023{\natexlab{b}})\citenamefont {Katsimiga}, \citenamefont {Mistakidis}, \citenamefont {Malomed}, \citenamefont {Frantzeskakis}, \citenamefont {Carretero-Gonzalez},\ and\ \citenamefont {Kevrekidis}}]{katsimiga2023interactions}%
  \BibitemOpen
  \bibfield  {author} {\bibinfo {author} {\bibfnamefont {G.~C.}\ \bibnamefont {Katsimiga}}, \bibinfo {author} {\bibfnamefont {S.~I.}\ \bibnamefont {Mistakidis}}, \bibinfo {author} {\bibfnamefont {B.~A.}\ \bibnamefont {Malomed}}, \bibinfo {author} {\bibfnamefont {D.~J.}\ \bibnamefont {Frantzeskakis}}, \bibinfo {author} {\bibfnamefont {R.}~\bibnamefont {Carretero-Gonzalez}}, \ and\ \bibinfo {author} {\bibfnamefont {P.~G.}\ \bibnamefont {Kevrekidis}},\ }\href {\doibase https://doi.org/10.3390/condmat8030067} {\bibfield  {journal} {\bibinfo  {journal} {Condensed Matter}\ }\textbf {\bibinfo {volume} {8}},\ \bibinfo {pages} {67} (\bibinfo {year} {2023}{\natexlab{b}})}\BibitemShut {NoStop}%
\bibitem [{\citenamefont {Tononi}\ \emph {et~al.}(2019)\citenamefont {Tononi}, \citenamefont {Wang},\ and\ \citenamefont {Salasnich}}]{tononi2019quantum}%
  \BibitemOpen
  \bibfield  {author} {\bibinfo {author} {\bibfnamefont {A.}~\bibnamefont {Tononi}}, \bibinfo {author} {\bibfnamefont {Y.}~\bibnamefont {Wang}}, \ and\ \bibinfo {author} {\bibfnamefont {L.}~\bibnamefont {Salasnich}},\ }\href {\doibase https://doi.org/10.1103/PhysRevA.99.063618} {\bibfield  {journal} {\bibinfo  {journal} {Phys. Rev. A}\ }\textbf {\bibinfo {volume} {99}},\ \bibinfo {pages} {063618} (\bibinfo {year} {2019})}\BibitemShut {NoStop}%
\bibitem [{\citenamefont {Cui}(2018)}]{PhysRevA.98.023630}%
  \BibitemOpen
  \bibfield  {author} {\bibinfo {author} {\bibfnamefont {X.}~\bibnamefont {Cui}},\ }\href {\doibase 10.1103/PhysRevA.98.023630} {\bibfield  {journal} {\bibinfo  {journal} {Phys. Rev. A}\ }\textbf {\bibinfo {volume} {98}},\ \bibinfo {pages} {023630} (\bibinfo {year} {2018})}\BibitemShut {NoStop}%
\bibitem [{\citenamefont {Gangwar}\ \emph {et~al.}(2024)\citenamefont {Gangwar}, \citenamefont {Ravisankar}, \citenamefont {Mistakidis}, \citenamefont {Muruganandam},\ and\ \citenamefont {Mishra}}]{gangwar2024spectrum}%
  \BibitemOpen
  \bibfield  {author} {\bibinfo {author} {\bibfnamefont {S.}~\bibnamefont {Gangwar}}, \bibinfo {author} {\bibfnamefont {R.}~\bibnamefont {Ravisankar}}, \bibinfo {author} {\bibfnamefont {S.~I.}\ \bibnamefont {Mistakidis}}, \bibinfo {author} {\bibfnamefont {P.}~\bibnamefont {Muruganandam}}, \ and\ \bibinfo {author} {\bibfnamefont {P.~K.}\ \bibnamefont {Mishra}},\ }\href {\doibase https://doi.org/10.1103/PhysRevA.109.013321} {\bibfield  {journal} {\bibinfo  {journal} {Phys. Rev. A}\ }\textbf {\bibinfo {volume} {109}},\ \bibinfo {pages} {013321} (\bibinfo {year} {2024})}\BibitemShut {NoStop}%
\bibitem [{\citenamefont {Rakshit}\ \emph {et~al.}(2019{\natexlab{a}})\citenamefont {Rakshit}, \citenamefont {Karpiuk}, \citenamefont {Brewczyk},\ and\ \citenamefont {Gajda}}]{rakshit2019quantum}%
  \BibitemOpen
  \bibfield  {author} {\bibinfo {author} {\bibfnamefont {D.}~\bibnamefont {Rakshit}}, \bibinfo {author} {\bibfnamefont {T.}~\bibnamefont {Karpiuk}}, \bibinfo {author} {\bibfnamefont {M.}~\bibnamefont {Brewczyk}}, \ and\ \bibinfo {author} {\bibfnamefont {M.}~\bibnamefont {Gajda}},\ }\href {\doibase doi: 10.21468/SciPostPhys.6.6.079} {\bibfield  {journal} {\bibinfo  {journal} {SciPost Phys.}\ }\textbf {\bibinfo {volume} {6}},\ \bibinfo {pages} {079} (\bibinfo {year} {2019}{\natexlab{a}})}\BibitemShut {NoStop}%
\bibitem [{\citenamefont {Rakshit}\ \emph {et~al.}(2019{\natexlab{b}})\citenamefont {Rakshit}, \citenamefont {Karpiuk}, \citenamefont {Zin}, \citenamefont {Brewczyk}, \citenamefont {Lewenstein},\ and\ \citenamefont {Gajda}}]{rakshit2019self}%
  \BibitemOpen
  \bibfield  {author} {\bibinfo {author} {\bibfnamefont {D.}~\bibnamefont {Rakshit}}, \bibinfo {author} {\bibfnamefont {T.}~\bibnamefont {Karpiuk}}, \bibinfo {author} {\bibfnamefont {P.}~\bibnamefont {Zin}}, \bibinfo {author} {\bibfnamefont {M.}~\bibnamefont {Brewczyk}}, \bibinfo {author} {\bibfnamefont {M.}~\bibnamefont {Lewenstein}}, \ and\ \bibinfo {author} {\bibfnamefont {M.}~\bibnamefont {Gajda}},\ }\href {\doibase 10.1088/1367-2630/ab2ce3} {\bibfield  {journal} {\bibinfo  {journal} {New J. Phys.}\ }\textbf {\bibinfo {volume} {21}},\ \bibinfo {pages} {073027} (\bibinfo {year} {2019}{\natexlab{b}})}\BibitemShut {NoStop}%
\bibitem [{\citenamefont {Salasnich}\ \emph {et~al.}(2007)\citenamefont {Salasnich}, \citenamefont {Adhikari},\ and\ \citenamefont {Toigo}}]{salasnich2007self}%
  \BibitemOpen
  \bibfield  {author} {\bibinfo {author} {\bibfnamefont {L.}~\bibnamefont {Salasnich}}, \bibinfo {author} {\bibfnamefont {S.~K.}\ \bibnamefont {Adhikari}}, \ and\ \bibinfo {author} {\bibfnamefont {F.}~\bibnamefont {Toigo}},\ }\href {\doibase https://doi.org/10.1103/PhysRevA.75.023616} {\bibfield  {journal} {\bibinfo  {journal} {Phys. Rev. A}\ }\textbf {\bibinfo {volume} {75}},\ \bibinfo {pages} {023616} (\bibinfo {year} {2007})}\BibitemShut {NoStop}%
\bibitem [{\citenamefont {Karpiuk}\ \emph {et~al.}(2004{\natexlab{a}})\citenamefont {Karpiuk}, \citenamefont {Brewczyk}, \citenamefont {Ospelkaus-Schwarzer}, \citenamefont {Bongs}, \citenamefont {Gajda},\ and\ \citenamefont {Rzą{\.z}ewski}}]{karpiuk2004soliton}%
  \BibitemOpen
  \bibfield  {author} {\bibinfo {author} {\bibfnamefont {T.}~\bibnamefont {Karpiuk}}, \bibinfo {author} {\bibfnamefont {M.}~\bibnamefont {Brewczyk}}, \bibinfo {author} {\bibfnamefont {S.}~\bibnamefont {Ospelkaus-Schwarzer}}, \bibinfo {author} {\bibfnamefont {K.}~\bibnamefont {Bongs}}, \bibinfo {author} {\bibfnamefont {M.}~\bibnamefont {Gajda}}, \ and\ \bibinfo {author} {\bibfnamefont {K.}~\bibnamefont {Rzą{\.z}ewski}},\ }\href {\doibase https://doi.org/10.1103/PhysRevLett.93.100401} {\bibfield  {journal} {\bibinfo  {journal} {Phys. Rev. Lett.}\ }\textbf {\bibinfo {volume} {93}},\ \bibinfo {pages} {100401} (\bibinfo {year} {2004}{\natexlab{a}})}\BibitemShut {NoStop}%
\bibitem [{\citenamefont {Karpiuk}\ \emph {et~al.}(2006)\citenamefont {Karpiuk}, \citenamefont {Brewczyk},\ and\ \citenamefont {Rzą{\.z}ewski}}]{karpiuk2006bright}%
  \BibitemOpen
  \bibfield  {author} {\bibinfo {author} {\bibfnamefont {T.}~\bibnamefont {Karpiuk}}, \bibinfo {author} {\bibfnamefont {M.}~\bibnamefont {Brewczyk}}, \ and\ \bibinfo {author} {\bibfnamefont {K.}~\bibnamefont {Rzą{\.z}ewski}},\ }\href {\doibase https://doi.org/10.1103/PhysRevA.73.053602} {\bibfield  {journal} {\bibinfo  {journal} {Phys. Rev. A}\ }\textbf {\bibinfo {volume} {73}},\ \bibinfo {pages} {053602} (\bibinfo {year} {2006})}\BibitemShut {NoStop}%
\bibitem [{\citenamefont {Massignan}\ \emph {et~al.}(2014)\citenamefont {Massignan}, \citenamefont {Zaccanti},\ and\ \citenamefont {Bruun}}]{massignan2014polarons}%
  \BibitemOpen
  \bibfield  {author} {\bibinfo {author} {\bibfnamefont {P.}~\bibnamefont {Massignan}}, \bibinfo {author} {\bibfnamefont {M.}~\bibnamefont {Zaccanti}}, \ and\ \bibinfo {author} {\bibfnamefont {G.~M.}\ \bibnamefont {Bruun}},\ }\href {\doibase 10.1088/0034-4885/77/3/034401} {\bibfield  {journal} {\bibinfo  {journal} {Rep. Progr. Phys.}\ }\textbf {\bibinfo {volume} {77}},\ \bibinfo {pages} {034401} (\bibinfo {year} {2014})}\BibitemShut {NoStop}%
\bibitem [{\citenamefont {Schmidt}\ \emph {et~al.}(2018)\citenamefont {Schmidt}, \citenamefont {Knap}, \citenamefont {Ivanov}, \citenamefont {You}, \citenamefont {Cetina},\ and\ \citenamefont {Demler}}]{schmidt2018universal}%
  \BibitemOpen
  \bibfield  {author} {\bibinfo {author} {\bibfnamefont {R.}~\bibnamefont {Schmidt}}, \bibinfo {author} {\bibfnamefont {M.}~\bibnamefont {Knap}}, \bibinfo {author} {\bibfnamefont {D.~A.}\ \bibnamefont {Ivanov}}, \bibinfo {author} {\bibfnamefont {J.-S.}\ \bibnamefont {You}}, \bibinfo {author} {\bibfnamefont {M.}~\bibnamefont {Cetina}}, \ and\ \bibinfo {author} {\bibfnamefont {E.}~\bibnamefont {Demler}},\ }\href {\doibase 10.1088/1361-6633/aa9593} {\bibfield  {journal} {\bibinfo  {journal} {Rep. Progr. Phys.}\ }\textbf {\bibinfo {volume} {81}},\ \bibinfo {pages} {024401} (\bibinfo {year} {2018})}\BibitemShut {NoStop}%
\bibitem [{\citenamefont {Mistakidis}\ \emph {et~al.}(2023)\citenamefont {Mistakidis}, \citenamefont {Volosniev}, \citenamefont {Barfknecht}, \citenamefont {Fogarty}, \citenamefont {Busch}, \citenamefont {Foerster}, \citenamefont {Schmelcher},\ and\ \citenamefont {Zinner}}]{mistakidis2023few}%
  \BibitemOpen
  \bibfield  {author} {\bibinfo {author} {\bibfnamefont {S.~I.}\ \bibnamefont {Mistakidis}}, \bibinfo {author} {\bibfnamefont {A.~G.}\ \bibnamefont {Volosniev}}, \bibinfo {author} {\bibfnamefont {R.~E.}\ \bibnamefont {Barfknecht}}, \bibinfo {author} {\bibfnamefont {T.}~\bibnamefont {Fogarty}}, \bibinfo {author} {\bibfnamefont {T.}~\bibnamefont {Busch}}, \bibinfo {author} {\bibfnamefont {A.}~\bibnamefont {Foerster}}, \bibinfo {author} {\bibfnamefont {P.}~\bibnamefont {Schmelcher}}, \ and\ \bibinfo {author} {\bibfnamefont {N.~T.}\ \bibnamefont {Zinner}},\ }\href {\doibase https://doi.org/10.1016/j.physrep.2023.10.004} {\bibfield  {journal} {\bibinfo  {journal} {Phys. Rep.}\ }\textbf {\bibinfo {volume} {1042}},\ \bibinfo {pages} {1} (\bibinfo {year} {2023})}\BibitemShut {NoStop}%
\bibitem [{\citenamefont {Abdullaev}\ and\ \citenamefont {Galimzyanov}(2020)}]{abdullaev2020bosonic}%
  \BibitemOpen
  \bibfield  {author} {\bibinfo {author} {\bibfnamefont {F.~K.}\ \bibnamefont {Abdullaev}}\ and\ \bibinfo {author} {\bibfnamefont {R.}~\bibnamefont {Galimzyanov}},\ }\href {\doibase 10.1088/1361-6455/ab9659} {\bibfield  {journal} {\bibinfo  {journal} {J. Phys. B: At.}\ }\textbf {\bibinfo {volume} {53}},\ \bibinfo {pages} {165301} (\bibinfo {year} {2020})}\BibitemShut {NoStop}%
\bibitem [{\citenamefont {Sinha}\ \emph {et~al.}(2023)\citenamefont {Sinha}, \citenamefont {Biswas}, \citenamefont {Santos},\ and\ \citenamefont {Sinha}}]{sinha2023impurities}%
  \BibitemOpen
  \bibfield  {author} {\bibinfo {author} {\bibfnamefont {S.}~\bibnamefont {Sinha}}, \bibinfo {author} {\bibfnamefont {S.}~\bibnamefont {Biswas}}, \bibinfo {author} {\bibfnamefont {L.}~\bibnamefont {Santos}}, \ and\ \bibinfo {author} {\bibfnamefont {S.}~\bibnamefont {Sinha}},\ }\href {\doibase 10.1103/PhysRevA.108.023311} {\bibfield  {journal} {\bibinfo  {journal} {Phys. Rev. A}\ }\textbf {\bibinfo {volume} {108}},\ \bibinfo {pages} {023311} (\bibinfo {year} {2023})}\BibitemShut {NoStop}%
\bibitem [{\citenamefont {Bighin}\ \emph {et~al.}(2022)\citenamefont {Bighin}, \citenamefont {Burchianti}, \citenamefont {Minardi},\ and\ \citenamefont {Macr{\`\i}}}]{bighin2022impurity}%
  \BibitemOpen
  \bibfield  {author} {\bibinfo {author} {\bibfnamefont {G.}~\bibnamefont {Bighin}}, \bibinfo {author} {\bibfnamefont {A.}~\bibnamefont {Burchianti}}, \bibinfo {author} {\bibfnamefont {F.}~\bibnamefont {Minardi}}, \ and\ \bibinfo {author} {\bibfnamefont {T.}~\bibnamefont {Macr{\`\i}}},\ }\href {\doibase https://doi.org/10.1103/PhysRevA.106.023301} {\bibfield  {journal} {\bibinfo  {journal} {Phys. Rev. A}\ }\textbf {\bibinfo {volume} {106}},\ \bibinfo {pages} {023301} (\bibinfo {year} {2022})}\BibitemShut {NoStop}%
\bibitem [{\citenamefont {Wenzel}\ \emph {et~al.}(2018)\citenamefont {Wenzel}, \citenamefont {Pfau},\ and\ \citenamefont {Ferrier-Barbut}}]{wenzel2018fermionic}%
  \BibitemOpen
  \bibfield  {author} {\bibinfo {author} {\bibfnamefont {M.}~\bibnamefont {Wenzel}}, \bibinfo {author} {\bibfnamefont {T.}~\bibnamefont {Pfau}}, \ and\ \bibinfo {author} {\bibfnamefont {I.}~\bibnamefont {Ferrier-Barbut}},\ }\href {\doibase 10.1088/1402-4896/aadd72} {\bibfield  {journal} {\bibinfo  {journal} {Phys. Scr.}\ }\textbf {\bibinfo {volume} {93}},\ \bibinfo {pages} {104004} (\bibinfo {year} {2018})}\BibitemShut {NoStop}%
\bibitem [{\citenamefont {Lous}\ \emph {et~al.}(2018)\citenamefont {Lous}, \citenamefont {Fritsche}, \citenamefont {Jag}, \citenamefont {Lehmann}, \citenamefont {Kirilov}, \citenamefont {Huang}, \citenamefont {Grimm} \emph {et~al.}}]{lous2018probing}%
  \BibitemOpen
  \bibfield  {author} {\bibinfo {author} {\bibfnamefont {R.~S.}\ \bibnamefont {Lous}}, \bibinfo {author} {\bibfnamefont {I.}~\bibnamefont {Fritsche}}, \bibinfo {author} {\bibfnamefont {M.}~\bibnamefont {Jag}}, \bibinfo {author} {\bibfnamefont {F.}~\bibnamefont {Lehmann}}, \bibinfo {author} {\bibfnamefont {E.}~\bibnamefont {Kirilov}}, \bibinfo {author} {\bibfnamefont {B.}~\bibnamefont {Huang}}, \bibinfo {author} {\bibfnamefont {R.}~\bibnamefont {Grimm}},  \emph {et~al.},\ }\href {\doibase https://doi.org/10.1103/PhysRevLett.120.243403} {\bibfield  {journal} {\bibinfo  {journal} {Phys. Rev. Lett.}\ }\textbf {\bibinfo {volume} {120}},\ \bibinfo {pages} {243403} (\bibinfo {year} {2018})}\BibitemShut {NoStop}%
\bibitem [{\citenamefont {Keller}\ \emph {et~al.}(2022)\citenamefont {Keller}, \citenamefont {Fogarty},\ and\ \citenamefont {Busch}}]{self-pinning}%
  \BibitemOpen
  \bibfield  {author} {\bibinfo {author} {\bibfnamefont {T.}~\bibnamefont {Keller}}, \bibinfo {author} {\bibfnamefont {T.}~\bibnamefont {Fogarty}}, \ and\ \bibinfo {author} {\bibfnamefont {T.}~\bibnamefont {Busch}},\ }\href {\doibase https://doi.org/10.1103/PhysRevLett.128.053401} {\bibfield  {journal} {\bibinfo  {journal} {Phys. Rev. Lett.}\ }\textbf {\bibinfo {volume} {128}},\ \bibinfo {pages} {053401} (\bibinfo {year} {2022})}\BibitemShut {NoStop}%
\bibitem [{\citenamefont {Keller}\ \emph {et~al.}(2023)\citenamefont {Keller}, \citenamefont {Fogarty},\ and\ \citenamefont {Busch}}]{Keller:23}%
  \BibitemOpen
  \bibfield  {author} {\bibinfo {author} {\bibfnamefont {T.}~\bibnamefont {Keller}}, \bibinfo {author} {\bibfnamefont {T.}~\bibnamefont {Fogarty}}, \ and\ \bibinfo {author} {\bibfnamefont {T.}~\bibnamefont {Busch}},\ }\href {\doibase 10.21468/SciPostPhys.15.3.095} {\bibfield  {journal} {\bibinfo  {journal} {SciPost Phys.}\ }\textbf {\bibinfo {volume} {15}},\ \bibinfo {pages} {095} (\bibinfo {year} {2023})}\BibitemShut {NoStop}%
\bibitem [{\citenamefont {Kwon}\ \emph {et~al.}(2021)\citenamefont {Kwon}, \citenamefont {Mukherjee}, \citenamefont {Huh}, \citenamefont {Kim}, \citenamefont {Mistakidis}, \citenamefont {Maity}, \citenamefont {Kevrekidis}, \citenamefont {Majumder}, \citenamefont {Schmelcher},\ and\ \citenamefont {Choi}}]{kwon2021spontaneous}%
  \BibitemOpen
  \bibfield  {author} {\bibinfo {author} {\bibfnamefont {K.}~\bibnamefont {Kwon}}, \bibinfo {author} {\bibfnamefont {K.}~\bibnamefont {Mukherjee}}, \bibinfo {author} {\bibfnamefont {S.~J.}\ \bibnamefont {Huh}}, \bibinfo {author} {\bibfnamefont {K.}~\bibnamefont {Kim}}, \bibinfo {author} {\bibfnamefont {S.~I.}\ \bibnamefont {Mistakidis}}, \bibinfo {author} {\bibfnamefont {D.~K.}\ \bibnamefont {Maity}}, \bibinfo {author} {\bibfnamefont {P.~G.}\ \bibnamefont {Kevrekidis}}, \bibinfo {author} {\bibfnamefont {S.}~\bibnamefont {Majumder}}, \bibinfo {author} {\bibfnamefont {P.}~\bibnamefont {Schmelcher}}, \ and\ \bibinfo {author} {\bibfnamefont {J.-y.}\ \bibnamefont {Choi}},\ }\href {\doibase https://doi.org/10.1103/PhysRevLett.127.113001} {\bibfield  {journal} {\bibinfo  {journal} {Phys. Rev. Lett.}\ }\textbf {\bibinfo {volume} {127}},\ \bibinfo {pages} {113001} (\bibinfo {year} {2021})}\BibitemShut {NoStop}%
\bibitem [{\citenamefont {Makhalov}\ \emph {et~al.}(2014)\citenamefont {Makhalov}, \citenamefont {Martiyanov},\ and\ \citenamefont {Turlapov}}]{makhalov2014ground}%
  \BibitemOpen
  \bibfield  {author} {\bibinfo {author} {\bibfnamefont {V.}~\bibnamefont {Makhalov}}, \bibinfo {author} {\bibfnamefont {K.}~\bibnamefont {Martiyanov}}, \ and\ \bibinfo {author} {\bibfnamefont {A.}~\bibnamefont {Turlapov}},\ }\href {\doibase https://doi.org/10.1103/PhysRevLett.112.045301} {\bibfield  {journal} {\bibinfo  {journal} {Phys. Rev. Lett.}\ }\textbf {\bibinfo {volume} {112}},\ \bibinfo {pages} {045301} (\bibinfo {year} {2014})}\BibitemShut {NoStop}%
\bibitem [{\citenamefont {Hu}\ \emph {et~al.}(2022)\citenamefont {Hu}, \citenamefont {Fei}, \citenamefont {Chen},\ and\ \citenamefont {Zhang}}]{hu2022collisional}%
  \BibitemOpen
  \bibfield  {author} {\bibinfo {author} {\bibfnamefont {Y.}~\bibnamefont {Hu}}, \bibinfo {author} {\bibfnamefont {Y.}~\bibnamefont {Fei}}, \bibinfo {author} {\bibfnamefont {X.-L.}\ \bibnamefont {Chen}}, \ and\ \bibinfo {author} {\bibfnamefont {Y.}~\bibnamefont {Zhang}},\ }\href {\doibase https://doi.org/10.1007/s11467-022-1233-7} {\bibfield  {journal} {\bibinfo  {journal} {Front. Phys.}\ }\textbf {\bibinfo {volume} {17}},\ \bibinfo {pages} {61505} (\bibinfo {year} {2022})}\BibitemShut {NoStop}%
\bibitem [{\citenamefont {Pethick}\ and\ \citenamefont {Smith}(2008)}]{pethick2008bose}%
  \BibitemOpen
  \bibfield  {author} {\bibinfo {author} {\bibfnamefont {C.~J.}\ \bibnamefont {Pethick}}\ and\ \bibinfo {author} {\bibfnamefont {H.}~\bibnamefont {Smith}},\ }\href {\doibase https://doi.org/10.1017/CBO9780511802850} {\emph {\bibinfo {title} {{B}ose-{E}instein condensation in dilute gases}}}\ (\bibinfo  {publisher} {Cambridge university press},\ \bibinfo {year} {2008})\BibitemShut {NoStop}%
\bibitem [{\citenamefont {Lewenstein}\ \emph {et~al.}(2012)\citenamefont {Lewenstein}, \citenamefont {Sanpera},\ and\ \citenamefont {Ahufinger}}]{lewenstein2012ultracold}%
  \BibitemOpen
  \bibfield  {author} {\bibinfo {author} {\bibfnamefont {M.}~\bibnamefont {Lewenstein}}, \bibinfo {author} {\bibfnamefont {A.}~\bibnamefont {Sanpera}}, \ and\ \bibinfo {author} {\bibfnamefont {V.}~\bibnamefont {Ahufinger}},\ }\href {\doibase https://doi.org/10.1093/acprof:oso/9780199573127.001.0001} {\emph {\bibinfo {title} {Ultracold Atoms in Optical Lattices: Simulating quantum many-body systems}}}\ (\bibinfo  {publisher} {OUP Oxford},\ \bibinfo {year} {2012})\BibitemShut {NoStop}%
\bibitem [{\citenamefont {Karpiuk}\ \emph {et~al.}(2004{\natexlab{b}})\citenamefont {Karpiuk}, \citenamefont {Brewczyk},\ and\ \citenamefont {Rz{\k{a}}{\.z}ewski}}]{karpiuk2004ground}%
  \BibitemOpen
  \bibfield  {author} {\bibinfo {author} {\bibfnamefont {T.}~\bibnamefont {Karpiuk}}, \bibinfo {author} {\bibfnamefont {M.}~\bibnamefont {Brewczyk}}, \ and\ \bibinfo {author} {\bibfnamefont {K.}~\bibnamefont {Rz{\k{a}}{\.z}ewski}},\ }\href {\doibase https://doi.org/10.1103/PhysRevA.69.043603} {\bibfield  {journal} {\bibinfo  {journal} {Phys. Rev. A}\ }\textbf {\bibinfo {volume} {69}},\ \bibinfo {pages} {043603} (\bibinfo {year} {2004}{\natexlab{b}})}\BibitemShut {NoStop}%
\bibitem [{\citenamefont {Weideman}\ and\ \citenamefont {Herbst}(1986)}]{weideman1986split}%
  \BibitemOpen
  \bibfield  {author} {\bibinfo {author} {\bibfnamefont {J.}~\bibnamefont {Weideman}}\ and\ \bibinfo {author} {\bibfnamefont {B.~M.}\ \bibnamefont {Herbst}},\ }\href {\doibase https://doi.org/10.1137/0723033} {\bibfield  {journal} {\bibinfo  {journal} {SIAM J. Numer. Anal.}\ }\textbf {\bibinfo {volume} {23}},\ \bibinfo {pages} {485} (\bibinfo {year} {1986})}\BibitemShut {NoStop}%
\bibitem [{\citenamefont {Gajda}\ \emph {et~al.}(2020)\citenamefont {Gajda}, \citenamefont {Mostowski}, \citenamefont {Pylak}, \citenamefont {Sowi{\'n}ski},\ and\ \citenamefont {Za{\l}uska-Kotur}}]{pauli_crystal2}%
  \BibitemOpen
  \bibfield  {author} {\bibinfo {author} {\bibfnamefont {M.}~\bibnamefont {Gajda}}, \bibinfo {author} {\bibfnamefont {J.}~\bibnamefont {Mostowski}}, \bibinfo {author} {\bibfnamefont {M.}~\bibnamefont {Pylak}}, \bibinfo {author} {\bibfnamefont {T.}~\bibnamefont {Sowi{\'n}ski}}, \ and\ \bibinfo {author} {\bibfnamefont {M.}~\bibnamefont {Za{\l}uska-Kotur}},\ }\href {\doibase https://doi.org/10.3390/sym12111886} {\bibfield  {journal} {\bibinfo  {journal} {Symmetry}\ }\textbf {\bibinfo {volume} {12}},\ \bibinfo {pages} {1886} (\bibinfo {year} {2020})}\BibitemShut {NoStop}%
\bibitem [{\citenamefont {Viverit}\ \emph {et~al.}(2000)\citenamefont {Viverit}, \citenamefont {Pethick},\ and\ \citenamefont {Smith}}]{viverit2000zero}%
  \BibitemOpen
  \bibfield  {author} {\bibinfo {author} {\bibfnamefont {L.}~\bibnamefont {Viverit}}, \bibinfo {author} {\bibfnamefont {C.~J.}\ \bibnamefont {Pethick}}, \ and\ \bibinfo {author} {\bibfnamefont {H.}~\bibnamefont {Smith}},\ }\href {\doibase https://doi.org/10.1103/PhysRevA.61.053605} {\bibfield  {journal} {\bibinfo  {journal} {Phys. Rev. A}\ }\textbf {\bibinfo {volume} {61}},\ \bibinfo {pages} {053605} (\bibinfo {year} {2000})}\BibitemShut {NoStop}%
\bibitem [{\citenamefont {Mistakidis}\ \emph {et~al.}(2019)\citenamefont {Mistakidis}, \citenamefont {Hilbig},\ and\ \citenamefont {Schmelcher}}]{mistakidis2019correlated}%
  \BibitemOpen
  \bibfield  {author} {\bibinfo {author} {\bibfnamefont {S.~I.}\ \bibnamefont {Mistakidis}}, \bibinfo {author} {\bibfnamefont {L.}~\bibnamefont {Hilbig}}, \ and\ \bibinfo {author} {\bibfnamefont {P.}~\bibnamefont {Schmelcher}},\ }\href {\doibase https://doi.org/10.1103/PhysRevA.100.023620} {\bibfield  {journal} {\bibinfo  {journal} {Phys. Rev. A}\ }\textbf {\bibinfo {volume} {100}},\ \bibinfo {pages} {023620} (\bibinfo {year} {2019})}\BibitemShut {NoStop}%
\bibitem [{\citenamefont {Mistakidis}\ \emph {et~al.}(2020{\natexlab{a}})\citenamefont {Mistakidis}, \citenamefont {Koutentakis}, \citenamefont {Katsimiga}, \citenamefont {Busch},\ and\ \citenamefont {Schmelcher}}]{distance_measure}%
  \BibitemOpen
  \bibfield  {author} {\bibinfo {author} {\bibfnamefont {S.~I.}\ \bibnamefont {Mistakidis}}, \bibinfo {author} {\bibfnamefont {G.~M.}\ \bibnamefont {Koutentakis}}, \bibinfo {author} {\bibfnamefont {G.~C.}\ \bibnamefont {Katsimiga}}, \bibinfo {author} {\bibfnamefont {T.}~\bibnamefont {Busch}}, \ and\ \bibinfo {author} {\bibfnamefont {P.}~\bibnamefont {Schmelcher}},\ }\href {\doibase 10.1088/1367-2630/ab7599} {\bibfield  {journal} {\bibinfo  {journal} {New J. Phys.}\ }\textbf {\bibinfo {volume} {22}},\ \bibinfo {pages} {043007} (\bibinfo {year} {2020}{\natexlab{a}})}\BibitemShut {NoStop}%
\bibitem [{\citenamefont {Keiler}\ \emph {et~al.}(2020)\citenamefont {Keiler}, \citenamefont {Mistakidis},\ and\ \citenamefont {Schmelcher}}]{distance_measure_1D}%
  \BibitemOpen
  \bibfield  {author} {\bibinfo {author} {\bibfnamefont {K.}~\bibnamefont {Keiler}}, \bibinfo {author} {\bibfnamefont {S.~I.}\ \bibnamefont {Mistakidis}}, \ and\ \bibinfo {author} {\bibfnamefont {P.}~\bibnamefont {Schmelcher}},\ }\href {\doibase 10.1088/1367-2630/ab9e34} {\bibfield  {journal} {\bibinfo  {journal} {New J. Phys.}\ }\textbf {\bibinfo {volume} {22}},\ \bibinfo {pages} {083003} (\bibinfo {year} {2020})}\BibitemShut {NoStop}%
\bibitem [{\citenamefont {Henseler}\ and\ \citenamefont {Shapiro}(2008)}]{density_correlations}%
  \BibitemOpen
  \bibfield  {author} {\bibinfo {author} {\bibfnamefont {P.}~\bibnamefont {Henseler}}\ and\ \bibinfo {author} {\bibfnamefont {B.}~\bibnamefont {Shapiro}},\ }\href {\doibase https://doi.org/10.1103/PhysRevA.77.033624} {\bibfield  {journal} {\bibinfo  {journal} {Phys. Rev. A}\ }\textbf {\bibinfo {volume} {77}},\ \bibinfo {pages} {033624} (\bibinfo {year} {2008})}\BibitemShut {NoStop}%
\bibitem [{\citenamefont {Bergschneider}\ \emph {et~al.}(2018)\citenamefont {Bergschneider}, \citenamefont {Klinkhamer}, \citenamefont {Becher}, \citenamefont {Klemt}, \citenamefont {Z{\"u}rn}, \citenamefont {Preiss},\ and\ \citenamefont {Jochim}}]{bergschneider2018spin}%
  \BibitemOpen
  \bibfield  {author} {\bibinfo {author} {\bibfnamefont {A.}~\bibnamefont {Bergschneider}}, \bibinfo {author} {\bibfnamefont {V.~M.}\ \bibnamefont {Klinkhamer}}, \bibinfo {author} {\bibfnamefont {J.~H.}\ \bibnamefont {Becher}}, \bibinfo {author} {\bibfnamefont {R.}~\bibnamefont {Klemt}}, \bibinfo {author} {\bibfnamefont {G.}~\bibnamefont {Z{\"u}rn}}, \bibinfo {author} {\bibfnamefont {P.~M.}\ \bibnamefont {Preiss}}, \ and\ \bibinfo {author} {\bibfnamefont {S.}~\bibnamefont {Jochim}},\ }\href {\doibase https://doi.org/10.1103/PhysRevA.97.063613} {\bibfield  {journal} {\bibinfo  {journal} {Phys. Rev. A}\ }\textbf {\bibinfo {volume} {97}},\ \bibinfo {pages} {063613} (\bibinfo {year} {2018})}\BibitemShut {NoStop}%
\bibitem [{\citenamefont {Will}\ \emph {et~al.}(2021)\citenamefont {Will}, \citenamefont {Astrakharchik},\ and\ \citenamefont {Fleischhauer}}]{will2021polaron}%
  \BibitemOpen
  \bibfield  {author} {\bibinfo {author} {\bibfnamefont {M.}~\bibnamefont {Will}}, \bibinfo {author} {\bibfnamefont {G.~E.}\ \bibnamefont {Astrakharchik}}, \ and\ \bibinfo {author} {\bibfnamefont {M.}~\bibnamefont {Fleischhauer}},\ }\href {\doibase https://doi.org/10.1103/PhysRevLett.127.103401} {\bibfield  {journal} {\bibinfo  {journal} {Phys. Rev. Lett.}\ }\textbf {\bibinfo {volume} {127}},\ \bibinfo {pages} {103401} (\bibinfo {year} {2021})}\BibitemShut {NoStop}%
\bibitem [{\citenamefont {Petkovi{\'c}}\ and\ \citenamefont {Ristivojevic}(2022)}]{petkovic2022mediated}%
  \BibitemOpen
  \bibfield  {author} {\bibinfo {author} {\bibfnamefont {A.}~\bibnamefont {Petkovi{\'c}}}\ and\ \bibinfo {author} {\bibfnamefont {Z.}~\bibnamefont {Ristivojevic}},\ }\href {\doibase https://doi.org/10.1103/PhysRevA.105.L021303} {\bibfield  {journal} {\bibinfo  {journal} {Phys. Rev. A}\ }\textbf {\bibinfo {volume} {105}},\ \bibinfo {pages} {L021303} (\bibinfo {year} {2022})}\BibitemShut {NoStop}%
\bibitem [{\citenamefont {Pasek}\ and\ \citenamefont {Orso}(2019)}]{pasek2019induced}%
  \BibitemOpen
  \bibfield  {author} {\bibinfo {author} {\bibfnamefont {M.}~\bibnamefont {Pasek}}\ and\ \bibinfo {author} {\bibfnamefont {G.}~\bibnamefont {Orso}},\ }\href {\doibase https://doi.org/10.1103/PhysRevB.100.245419} {\bibfield  {journal} {\bibinfo  {journal} {Phys. Rev. B}\ }\textbf {\bibinfo {volume} {100}},\ \bibinfo {pages} {245419} (\bibinfo {year} {2019})}\BibitemShut {NoStop}%
\bibitem [{\citenamefont {Baroni}\ \emph {et~al.}(2024)\citenamefont {Baroni}, \citenamefont {Huang}, \citenamefont {Fritsche}, \citenamefont {Dobler}, \citenamefont {Anich}, \citenamefont {Kirilov}, \citenamefont {Grimm}, \citenamefont {Bastarrachea-Magnani}, \citenamefont {Massignan},\ and\ \citenamefont {Bruun}}]{baroni2024mediated}%
  \BibitemOpen
  \bibfield  {author} {\bibinfo {author} {\bibfnamefont {C.}~\bibnamefont {Baroni}}, \bibinfo {author} {\bibfnamefont {B.}~\bibnamefont {Huang}}, \bibinfo {author} {\bibfnamefont {I.}~\bibnamefont {Fritsche}}, \bibinfo {author} {\bibfnamefont {E.}~\bibnamefont {Dobler}}, \bibinfo {author} {\bibfnamefont {G.}~\bibnamefont {Anich}}, \bibinfo {author} {\bibfnamefont {E.}~\bibnamefont {Kirilov}}, \bibinfo {author} {\bibfnamefont {R.}~\bibnamefont {Grimm}}, \bibinfo {author} {\bibfnamefont {M.~A.}\ \bibnamefont {Bastarrachea-Magnani}}, \bibinfo {author} {\bibfnamefont {P.}~\bibnamefont {Massignan}}, \ and\ \bibinfo {author} {\bibfnamefont {G.~M.}\ \bibnamefont {Bruun}},\ }\href {\doibase https://doi.org/10.1038/s41567-023-02248-4} {\bibfield  {journal} {\bibinfo  {journal} {Nat. Phys.}\ }\textbf {\bibinfo {volume} {20}},\ \bibinfo {pages} {68} (\bibinfo {year} {2024})}\BibitemShut {NoStop}%
\bibitem [{\citenamefont {Gampel}\ \emph {et~al.}(2022)\citenamefont {Gampel}, \citenamefont {Gajda}, \citenamefont {Za{\l}uska-Kotur},\ and\ \citenamefont {Mostowski}}]{pauli_crystal1}%
  \BibitemOpen
  \bibfield  {author} {\bibinfo {author} {\bibfnamefont {F.}~\bibnamefont {Gampel}}, \bibinfo {author} {\bibfnamefont {M.}~\bibnamefont {Gajda}}, \bibinfo {author} {\bibfnamefont {M.}~\bibnamefont {Za{\l}uska-Kotur}}, \ and\ \bibinfo {author} {\bibfnamefont {J.}~\bibnamefont {Mostowski}},\ }\href {\doibase https://doi.org/10.1016/j.physleta.2021.127799} {\bibfield  {journal} {\bibinfo  {journal} {Phys. Lett. A}\ }\textbf {\bibinfo {volume} {422}},\ \bibinfo {pages} {127799} (\bibinfo {year} {2022})}\BibitemShut {NoStop}%
\bibitem [{\citenamefont {Kevrekidis}\ \emph {et~al.}(2008)\citenamefont {Kevrekidis}, \citenamefont {Frantzeskakis},\ and\ \citenamefont {Carretero-Gonz{\'a}lez}}]{kevrekidis2008emergent}%
  \BibitemOpen
  \bibfield  {author} {\bibinfo {author} {\bibfnamefont {P.~G.}\ \bibnamefont {Kevrekidis}}, \bibinfo {author} {\bibfnamefont {D.~J.}\ \bibnamefont {Frantzeskakis}}, \ and\ \bibinfo {author} {\bibfnamefont {R.}~\bibnamefont {Carretero-Gonz{\'a}lez}},\ }\href {\doibase https://doi.org/10.1007/978-3-540-73591-5} {\emph {\bibinfo {title} {Emergent nonlinear phenomena in Bose-Einstein condensates: theory and experiment}}},\ Vol.~\bibinfo {volume} {45}\ (\bibinfo  {publisher} {Springer},\ \bibinfo {year} {2008})\BibitemShut {NoStop}%
\bibitem [{\citenamefont {Jain}\ and\ \citenamefont {Boninsegni}(2011)}]{jain2011quantum}%
  \BibitemOpen
  \bibfield  {author} {\bibinfo {author} {\bibfnamefont {P.}~\bibnamefont {Jain}}\ and\ \bibinfo {author} {\bibfnamefont {M.}~\bibnamefont {Boninsegni}},\ }\href {\doibase https://doi.org/10.1103/PhysRevA.83.023602} {\bibfield  {journal} {\bibinfo  {journal} {Phys. Rev. A}\ }\textbf {\bibinfo {volume} {83}},\ \bibinfo {pages} {023602} (\bibinfo {year} {2011})}\BibitemShut {NoStop}%
\bibitem [{\citenamefont {Mistakidis}\ \emph {et~al.}(2018)\citenamefont {Mistakidis}, \citenamefont {Katsimiga}, \citenamefont {Kevrekidis},\ and\ \citenamefont {Schmelcher}}]{mistakidis2018correlation}%
  \BibitemOpen
  \bibfield  {author} {\bibinfo {author} {\bibfnamefont {S.~I.}\ \bibnamefont {Mistakidis}}, \bibinfo {author} {\bibfnamefont {G.~C.}\ \bibnamefont {Katsimiga}}, \bibinfo {author} {\bibfnamefont {P.~G.}\ \bibnamefont {Kevrekidis}}, \ and\ \bibinfo {author} {\bibfnamefont {P.}~\bibnamefont {Schmelcher}},\ }\href {\doibase 10.1088/1367-2630/aabc6a} {\bibfield  {journal} {\bibinfo  {journal} {New J. Phys.}\ }\textbf {\bibinfo {volume} {20}},\ \bibinfo {pages} {043052} (\bibinfo {year} {2018})}\BibitemShut {NoStop}%
\bibitem [{\citenamefont {Fogarty}\ \emph {et~al.}(2020)\citenamefont {Fogarty}, \citenamefont {Deffner}, \citenamefont {Busch},\ and\ \citenamefont {Campbell}}]{fogarty2020orthogonality}%
  \BibitemOpen
  \bibfield  {author} {\bibinfo {author} {\bibfnamefont {T.}~\bibnamefont {Fogarty}}, \bibinfo {author} {\bibfnamefont {S.}~\bibnamefont {Deffner}}, \bibinfo {author} {\bibfnamefont {T.}~\bibnamefont {Busch}}, \ and\ \bibinfo {author} {\bibfnamefont {S.}~\bibnamefont {Campbell}},\ }\href {\doibase https://doi.org/10.1103/PhysRevLett.124.110601} {\bibfield  {journal} {\bibinfo  {journal} {Phys. Rev. Lett.}\ }\textbf {\bibinfo {volume} {124}},\ \bibinfo {pages} {110601} (\bibinfo {year} {2020})}\BibitemShut {NoStop}%
\bibitem [{\citenamefont {Goold}\ \emph {et~al.}(2011)\citenamefont {Goold}, \citenamefont {Fogarty}, \citenamefont {Gullo}, \citenamefont {Paternostro},\ and\ \citenamefont {Busch}}]{goold2011orthogonality}%
  \BibitemOpen
  \bibfield  {author} {\bibinfo {author} {\bibfnamefont {J.}~\bibnamefont {Goold}}, \bibinfo {author} {\bibfnamefont {T.}~\bibnamefont {Fogarty}}, \bibinfo {author} {\bibfnamefont {N.~L.}\ \bibnamefont {Gullo}}, \bibinfo {author} {\bibfnamefont {M.}~\bibnamefont {Paternostro}}, \ and\ \bibinfo {author} {\bibfnamefont {T.}~\bibnamefont {Busch}},\ }\href {\doibase https://doi.org/10.1103/PhysRevA.84.063632} {\bibfield  {journal} {\bibinfo  {journal} {Phys. Rev. A}\ }\textbf {\bibinfo {volume} {84}},\ \bibinfo {pages} {063632} (\bibinfo {year} {2011})}\BibitemShut {NoStop}%
\bibitem [{\citenamefont {Otajonov}\ \emph {et~al.}(2022)\citenamefont {Otajonov}, \citenamefont {Tsoy},\ and\ \citenamefont {Abdullaev}}]{otajonov2022modulational}%
  \BibitemOpen
  \bibfield  {author} {\bibinfo {author} {\bibfnamefont {S.~R.}\ \bibnamefont {Otajonov}}, \bibinfo {author} {\bibfnamefont {E.~N.}\ \bibnamefont {Tsoy}}, \ and\ \bibinfo {author} {\bibfnamefont {F.~K.}\ \bibnamefont {Abdullaev}},\ }\href {\doibase https://doi.org/10.1103/PhysRevA.106.033309} {\bibfield  {journal} {\bibinfo  {journal} {Phys. Rev. A}\ }\textbf {\bibinfo {volume} {106}},\ \bibinfo {pages} {033309} (\bibinfo {year} {2022})}\BibitemShut {NoStop}%
\bibitem [{\citenamefont {Mithun}\ \emph {et~al.}(2020)\citenamefont {Mithun}, \citenamefont {Maluckov}, \citenamefont {Kasamatsu}, \citenamefont {Malomed},\ and\ \citenamefont {Khare}}]{mithun2020modulational}%
  \BibitemOpen
  \bibfield  {author} {\bibinfo {author} {\bibfnamefont {T.}~\bibnamefont {Mithun}}, \bibinfo {author} {\bibfnamefont {A.}~\bibnamefont {Maluckov}}, \bibinfo {author} {\bibfnamefont {K.}~\bibnamefont {Kasamatsu}}, \bibinfo {author} {\bibfnamefont {B.~A.}\ \bibnamefont {Malomed}}, \ and\ \bibinfo {author} {\bibfnamefont {A.}~\bibnamefont {Khare}},\ }\href {\doibase https://doi.org/10.3390/sym12010174} {\bibfield  {journal} {\bibinfo  {journal} {Symmetry}\ }\textbf {\bibinfo {volume} {12}},\ \bibinfo {pages} {174} (\bibinfo {year} {2020})}\BibitemShut {NoStop}%
\bibitem [{\citenamefont {Mistakidis}\ \emph {et~al.}(2020{\natexlab{b}})\citenamefont {Mistakidis}, \citenamefont {Volosniev},\ and\ \citenamefont {Schmelcher}}]{mistakidis2020induced}%
  \BibitemOpen
  \bibfield  {author} {\bibinfo {author} {\bibfnamefont {S.~I.}\ \bibnamefont {Mistakidis}}, \bibinfo {author} {\bibfnamefont {A.~G.}\ \bibnamefont {Volosniev}}, \ and\ \bibinfo {author} {\bibfnamefont {P.}~\bibnamefont {Schmelcher}},\ }\href {\doibase https://doi.org/10.1103/PhysRevResearch.2.023154} {\bibfield  {journal} {\bibinfo  {journal} {Phys. Rev. Res.}\ }\textbf {\bibinfo {volume} {2}},\ \bibinfo {pages} {023154} (\bibinfo {year} {2020}{\natexlab{b}})}\BibitemShut {NoStop}%
\end{thebibliography}%

\end{document}